\documentclass[journal=jacsat,manuscript=article]{achemso} 

\usepackage{chemformula} 
\usepackage[T1]{fontenc} 
\usepackage{booktabs}
\usepackage{multirow}
\usepackage{siunitx}
\usepackage{caption}
\usepackage{threeparttable}
\usepackage{array}
\usepackage{amsmath}
\usepackage{amsfonts}
\usepackage{hyperref}
\usepackage{graphicx}
\usepackage{subcaption} 
\usepackage[export]{adjustbox}
\usepackage{dcolumn}
\usepackage{bm}
\usepackage{braket}
\usepackage{ulem}
\usepackage{color}
\usepackage{float}
\usepackage{natbib}
\usepackage{xpatch}

\hypersetup{colorlinks=true, citecolor=blue, urlcolor=blue, linkcolor=blue}

\author{Xinyu Xu}
\affiliation{School of Physics, Anhui University, Hefei 230601, China}

\author{Rajibul Islam}
\affiliation{Department of Physics, University of Alabama at Birmingham, Birmingham 35294, AL, USA}

\author{Ghulam Hussain}
\affiliation{Institute for Advanced Study, Shenzhen University, Shenzhen 518060, China}

\author{Yangming Huang}
\affiliation{State Key Laboratory of Physical Chemistry of Solid Surfaces, Department of Chemistry, College of Chemistry and Chemical Engineering, and Fujian Provincial Key Laboratory of Theoretical and Computational Chemistry, Xiamen University, Xiamen 361005, China}

\author{Xiaoguang Li}
\affiliation{Institute for Advanced Study, Shenzhen University, Shenzhen 518060, China}

\author{Pavlo O. Dral}
\affiliation{State Key Laboratory of Physical Chemistry of Solid Surfaces, Department of Chemistry, College of Chemistry and Chemical Engineering, and Fujian Provincial Key Laboratory of Theoretical and Computational Chemistry, Xiamen University, Xiamen 361005, China}
\alsoaffiliation{Institute of Physics, Faculty of Physics, Astronomy, and Informatics, Nicolaus Copernicus University in Toru\'n, ul. Grudzi\k{a}dzka 5, 87-100 Toru\'n, Poland}
\alsoaffiliation{Institute of Advanced Studies, Nicolaus Copernicus University in Toru\'n, ul. Wile\'nska 4, 87-100 Toru\'n, Poland}
\alsoaffiliation{Atomistic, Shenzhen 518000, China}

\author{Arif Ullah}
\email{arif@ahu.edu.cn}
\affiliation{School of Physics, Anhui University, Hefei 230601, China}

\author{Ming Yang}
\email{mingyang@ahu.edu.cn}
\affiliation{School of Physics, Anhui University, Hefei 230601, China}

\title{Quantum-inspired Chemical Rule for Discovering Topological Materials}

\abbreviations{}

\begin{document}


\begin{abstract}
Topological materials exhibit unique electronic structures that underpin both fundamental quantum phenomena and next-generation technologies, yet their discovery remains constrained by the high computational cost of first-principles calculations and the slow, resource-intensive nature of experimental synthesis. Recent machine-learning approaches, such as the heuristic topogivity rule, offer a data-driven pre-screening tool by quantifying each element’s intrinsic tendency toward topological behavior. Here, we develop a hybrid quantum-classical neural network (HQCNN) that extends this rule into a quantum-inspired formulation. Within this framework, the HQCNN maps compositional descriptors to quantum probability amplitudes, naturally introducing pairwise inter-element correlations inaccessible to classical heuristics. The physical validity of these correlations is substantiated by constructing an equivalent complex-valued neural network (CVNN), confirming both the consistency and interpretability of the formulation. Retaining the simplicity of chemical reasoning while embedding quantum-native features, our quantum-inspired rule enables efficient and generalizable topological classification. High-throughput screening combined with first-principles (DFT) validation reveals five previously unreported topological compounds, demonstrating the enhanced predictive power and physical insight afforded by quantum-inspired heuristics.

\vspace{20pt}

\includegraphics[width=0.9\linewidth]{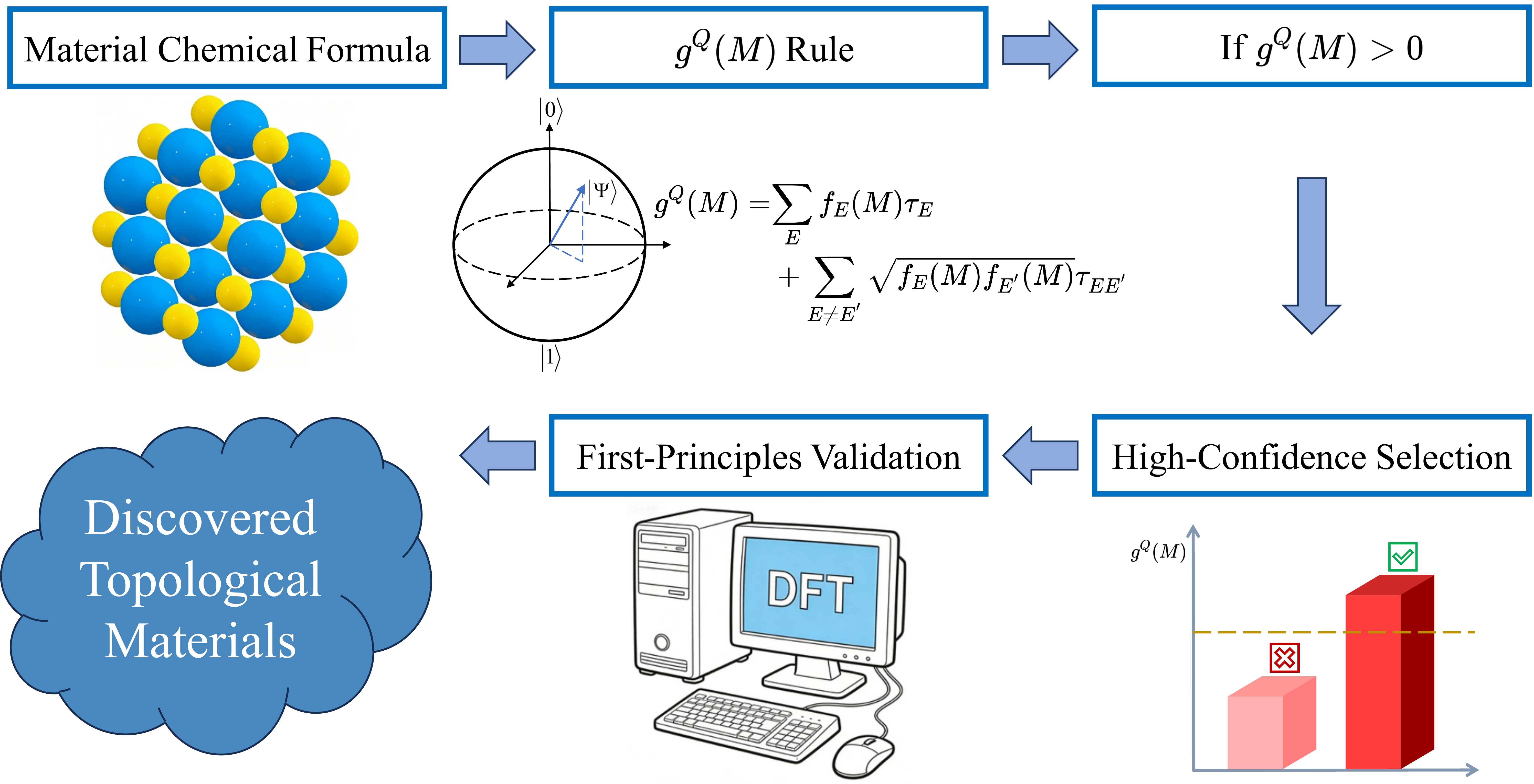}

\end{abstract}


\maketitle


\section{Introduction}

Topological materials are a vast category of quantum matter where the electronic wave functions possess nontrivial topological structures defined by quantized invariants.\cite{Haldane1,Kane2,Hasan3,Qi4,Armitage5} This family includes topological insulators (TIs), which feature an insulating interior but symmetry-protected metallic surface states,\cite{Kane6,Bernevig7,Konig8,Fu9,Hsieh10,Fu11,Hsieh12,Benalcazar13} and topological semimetals (TSMs), which host distinctive Weyl or Dirac nodes characterized by Berry-curvature textures.\cite{Liu14,Xu15,Lv16,Burkov17,Bradlyn18} The robust boundary modes in these materials lead to hallmark quantum responses—such as the quantum spin Hall effect\cite{Kane6} and anomalous transport\cite{Qi4}—that are highly promising for next-generation nanoscale quantum and spintronic technologies. To fully exploit the potential of these functional quantum materials, the primary obstacle is the development of efficient and universal criteria for reliably identifying whether a given compound has a topologically nontrivial electronic structure.

Traditional discovery paradigms rely on the synergy between first-principles electronic structure calculations and topological band theory, where topological invariants are directly computed from Bloch wave functions.\cite{Xiao19,Bansil20} Over the past decade, this foundation has been systematized through the development of topological quantum chemistry \cite{Bradlyn21} and symmetry indicator approaches.\cite{slager2013space, Po22,Fu23,Krutho24,Song25} These frameworks link symmetry eigenvalues at high-symmetry points to the global topology of electronic bands, thereby enabling large-scale, high-throughput screening of candidate topological materials.\cite{Tang26,Zhang27,Vergniory28} Their success has established symmetry-based methods as indispensable tools in modern topological materials discovery.

However, symmetry-based diagnostics are inherently incomplete. They are limited to crystalline systems where symmetry representations are well defined and cannot universally capture topological phases arising from accidental band inversions, low-symmetry distortions, or strong spin-orbit interactions beyond perturbative regimes. A paradigmatic example is the Weyl semimetal TaAs,\cite{Xu15,Lv16} whose topological character cannot be inferred from symmetry indicators alone but requires explicit computation of Berry curvature and Chern numbers from wave functions.\cite{Weng29} Similar challenges arise for Chern insulators lacking point-group symmetry and time-reversal-invariant $Z_2$ topological insulators, where direct evaluation of topological invariants remains indispensable. These limitations underscore a fundamental bottleneck: the absence of a unified and generalizable framework capable of identifying topological character across materials of arbitrary symmetry, chemical complexity, and correlation strength. 

{\color{black}In the past decade, the rapid advancement of machine learning has introduced a powerful new paradigm for materials discovery, including the prediction of topological phases \cite{Claussen31, Cao32, Schleder34, Andrejevic35, He2025, Xu2025predicting} and topological invariants \cite{Zhang38, Zhang39, Scheurer40} using various machine learning approaches based on diverse material representations such as electronic band structures, elemental properties, and symmetry information. Many of these approaches employ complex machine learning models (e.g., neural networks, gradient boosted trees, etc.). Although these models can achieve high predictive accuracy, they typically involve high-dimensional feature spaces and complex nonlinear transformations, making their internal decision-making processes relatively difficult to interpret. Even when feature importance analysis provides some degree of interpretability, extracting clear physical intuition from numerous complex features remains challenging.

In this context, researchers have also explored more interpretable approaches. For instance, Cao et al. \cite{Cao32} and Schleder et al. \cite{Schleder34} employed the sure independence screening and sparsifying operator (SISSO) method, a compressed sensing approach that automatically derives low-dimensional, analytically expressible descriptors from data, enabling interpretable predictions of topological materials. Subsequently, Ma et al. \cite{Ma36} proposed a particularly elegant approach that parallels the heuristic reasoning of chemistry: a data-driven chemical rule for diagnosing electronic topology that operates independently of crystalline symmetry. In contrast to relying on complex models, their method introduces a machine-learned elemental descriptor—termed topogivity—which quantifies each element's intrinsic tendency to participate in topological phases. The topological nature of a compound can then be inferred from the sign of the composition-weighted average of its constituent elements' topogivities, providing a simple yet remarkably effective heuristic that bridges atomic chemistry and topological quantum physics. The effectiveness of this rule has since been validated and modestly extended in subsequent studies. \cite{Xu37,Ma2025,Ullah2025TXL}}

\begin{figure*}[htbp]
\includegraphics[width=\textwidth]{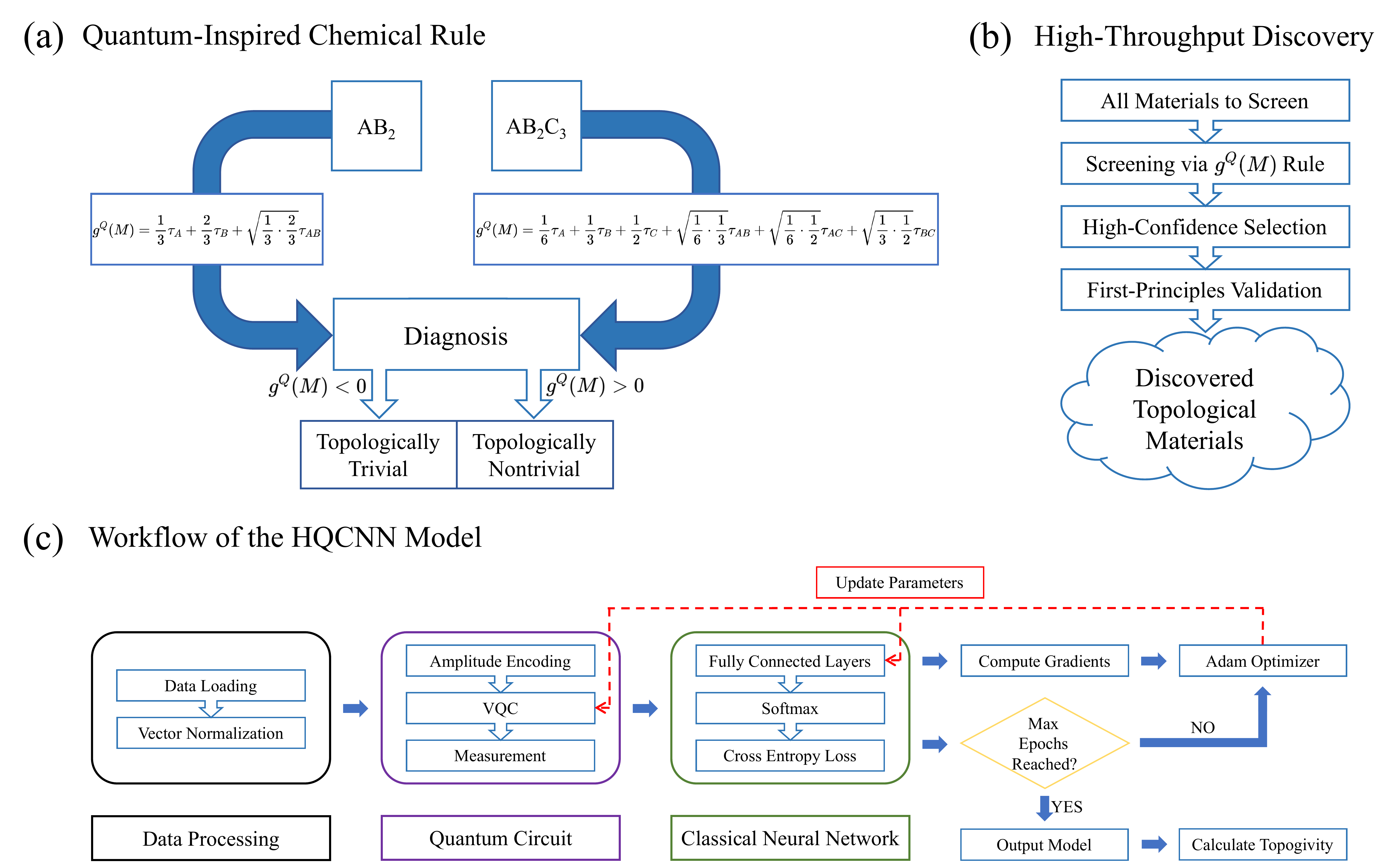} 
\caption{HQCNN-based diagnosis and discovery of topological materials. \textbf{(a)} Schematic illustration of the quantum-inspired heuristic rule $ g^{Q}(M) $ for topological diagnosis. For a given compound (e.g., $AB_{2}$ or $AB_{2}C_{3}$), the diagnostic value is obtained in two steps: first, by computing the weighted average of the constituent elements’ topogivities ($\tau_{A}, \tau_{B}, \tau_{C}$) according to their stoichiometric ratios; and second, by evaluating the weighted average of pairwise quantum topogivities ($\tau_{AB}, \tau_{AC}, \tau_{BC}$) between distinct elements, with weights given by the product of their relative abundances. The overall diagnostic value—sum of the two contributions—encodes both the topological classification (sign) and the confidence level (magnitude). \textbf{(b)} High-throughput discovery workflow using quantum-inspired chemical rule. Chemical formulae are first used for topological property prediction, followed by screening of high-confidence candidates. The shortlisted materials are then validated via DFT calculations, leading to the identification of previously unreported topological materials. \textbf{(c)} The general workflow of our HQCNN model, which integrates data preprocessing, a quantum circuit, and a classical fully-connected artificial neural network. Further details are given in Section~S1 of the Supporting Information (SI).}
  \label{fig:main_figure}
\end{figure*}
\par

In recent years, quantum machine learning (QML) has emerged as a powerful paradigm, integrating concepts from quantum computation and classical data-driven learning.\cite{Biamonte17,Schuld18} Within this framework, hybrid quantum-classical neural networks have become a standard architecture in the noisy intermediate-scale quantum (NISQ) era: a parameterized variationalquantum circuit (VQC) encodes classical data into quantum states, some or all qubits are measured, and the results feed into a classical neural network, with both quantum and classical parameters jointly optimized via backpropagation.\cite{chen2019vqnet,maleki2026qnas}{\color{black}In such architectures, the variational quantum circuit plays a role analogous to a hidden layer in a classical neural network, actively transforming the input features through trainable quantum gates.} Such hybrid models can capture entanglement and coherence effects beyond classical neural networks, and have been successfully applied to image classification tasks,\cite{Cheng2024_hybrid,basilewitsch2025quantum} establishing their viability for practical machine learning applications.

In materials physics and chemistry, many emergent properties arise not merely from constituent elements, but from intricate interactions between them—such as orbital hybridization in intermetallic compounds\cite{wang2024intermetallic} and charge transfer at interfaces.\cite{Jiang2024Charges} Capturing such inter-element correlations is essential for predicting material properties, yet remains challenging for classical heuristics that treat elements independently. This challenge is ideally suited to quantum-native representations, where probabilities from quantum measurement naturally encode pairwise interactions.

Motivated by these advances, in this work, we introduce a quantum-inspired chemical rule, $g^{Q}(M)$, derived from a hybrid quantum-classical neural network (HQCNN) (Figure~\ref{fig:main_figure}c)  designed to incorporate quantum advantages into the identification of topological materials. Here, "quantum-inspired" refers to the modeling philosophy: we utilize quantum mechanical concepts—such as probability amplitudes and interference—to naturally capture inter-element interactions. It is emphasized that this does not imply a quantum advantage in computation, and the HQCNN is implemented on classical computers via simulating quantum circuits.

Unlike the classical heuristic of Ma et al.,\cite{Ma36} which assigns an independent scalar propensity to each element, $g^{Q}(M)$ embeds pairwise correlation terms that encode inter-element quantum interactions absent in conventional models. The structure of these correlations is further validated using an equivalent complex-valued neural network (CVNN), which independently reproduces the same analytical form and reinforces the interpretability of the rule.

For practical screening, a material is classified as topological when $g^{Q}(M) > 0$ and trivial when $g^{Q}(M) < 0$ (Figure~\ref{fig:main_figure}a). Applying this rule in a high-throughput workflow, we screened a large materials database and performed first-principles (DFT) verification of representative predictions (Figure~\ref{fig:main_figure}b). Importantly, this quantum-inspired approach identifies five previously unreported topological materials overlooked by the classical counterpart, demonstrating its enhanced discovery capability and value for accelerating materials design.

\begin{figure*}
    \centering
    \includegraphics[width=\linewidth]{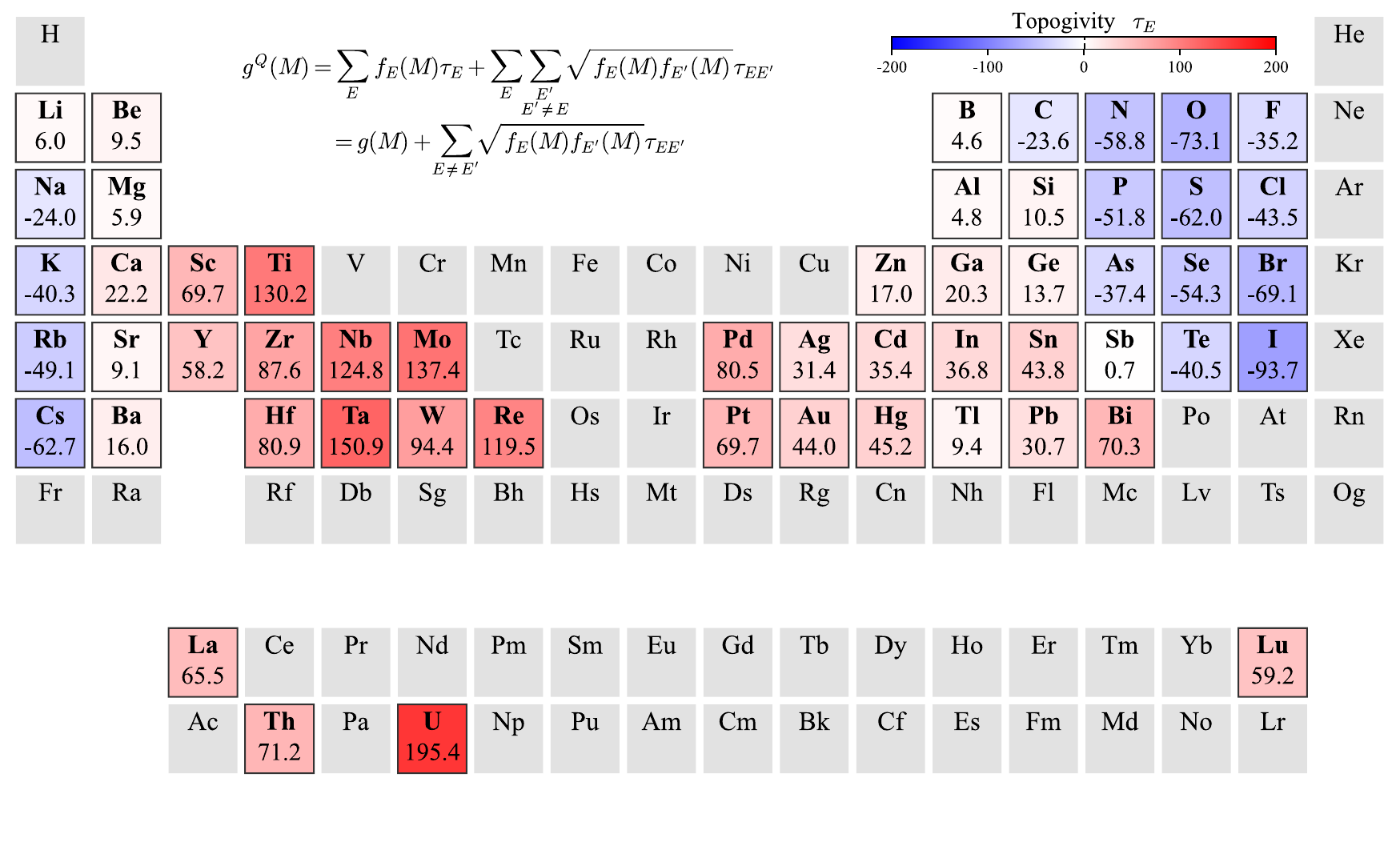}
    \caption{\color{black}{Periodic table of topogivities. Color-coding and numerical values represent HQCNN topogivities $\tau_{\rm E}$. Gray elements indicate those not present in the dataset. The learned values show chemically intuitive trends: heavy spin--orbit-coupled elements and early transition metals tend to have positive topogivities, whereas light elements, nonmetals, and halogens are weak or negative. The complete model contains 1,485 topogivity values (54 elemental topogivities shown here plus 1,431 cross-correlation terms not included in the periodic-table representation).}}
    \label{fig:ptable}
\end{figure*}

\section{Quantum-inspired Chemical Rule}

We begin by introducing the central outcome of this study—the quantum-inspired chemical rule $g^Q(M)$, expressed as:

\begin{align} \label{eq:gm}
g^Q(M)&= {\sum_{E} {f_{E}(M)}}\tau _{E}+{\sum_{E}\!\sum_{\substack{E^{'} \\ E^{'} \neq E}} \!{\textstyle\sqrt{f_{E}(M)f_{E^{'} }(M) } }\tau _{EE^{'} }}\\
    &=g(M)+ \!{\sum_{\substack{E \neq E^{'}} } \!{\textstyle\sqrt{f_{E}(M)f_{E^{'} }(M) } }\tau _{EE^{'}}} \, ,
\end{align}
where the first sum runs over all elements $E$ appearing in the chemical formula of material $M$, and the double sum runs over all distinct pairs of elements, with $E$ and $E'$ ordered such that the atomic number of $E$ is less than that of $E'$, and $f_{E}(M)$ and $f_{E'}(M)$ denote the fractional abundances of elements $E$ and $E'$, respectively.
The parameters $\tau_{E}$ represent element-specific contributions (the topogivity of $E$), while
$\tau_{EE^{'}}$ capture pairwise interactions between distinct elements $E$ and $E^{'}$.
For example, in a ternary compound $A_xB_yC_z$, the elemental fractions are
$f_{A}(M)=\frac{x}{x+y+z}$, $f_{B}(M)=\frac{y}{x+y+z}$, and $f_{C}(M)=\frac{z}{x+y+z}$.
In this case, the rule depends on both the individual topogivities $\tau_{A}$, $\tau_{B}$, and $\tau_{C}$, as well as the interaction parameters $\tau_{AB}$, $\tau_{AC}$, and $\tau_{BC}$, as illustrated in Figure~\ref{fig:main_figure}. The full derivation of $g^Q(M)$ rule is provided in Section~S1 of the Supporting Information (SI), and its complex-valued-network counterpart in Section~S2.

In Eq.~\eqref{eq:gm}, the first term recovers the purely classical rule proposed by Ma et al.\cite{Ma36}, while the second term represents the additional contribution originating from the quantum part of our HQCNN model. The classification decision is determined by the sign of $g^Q(M)$. If $g^Q(M)$ is positive, the material is classified as topologically nontrivial. If $g^Q(M)$ is negative, it is classified as topologically trivial. A larger absolute value of $g^Q(M)$ generally indicates a higher confidence level in the classification judgment. 

\section{Results and Discussion}

We adopt a supervised learning approach to derive chemically inspired rules for diagnosing topological materials. Our training dataset is taken from Ma et al.\cite{Ma36}, which itself was curated from a subset of the database compiled by Tang et al.\cite{Tang26}.

\begin{figure*}
    \centering
    \includegraphics[width=1.0\linewidth]{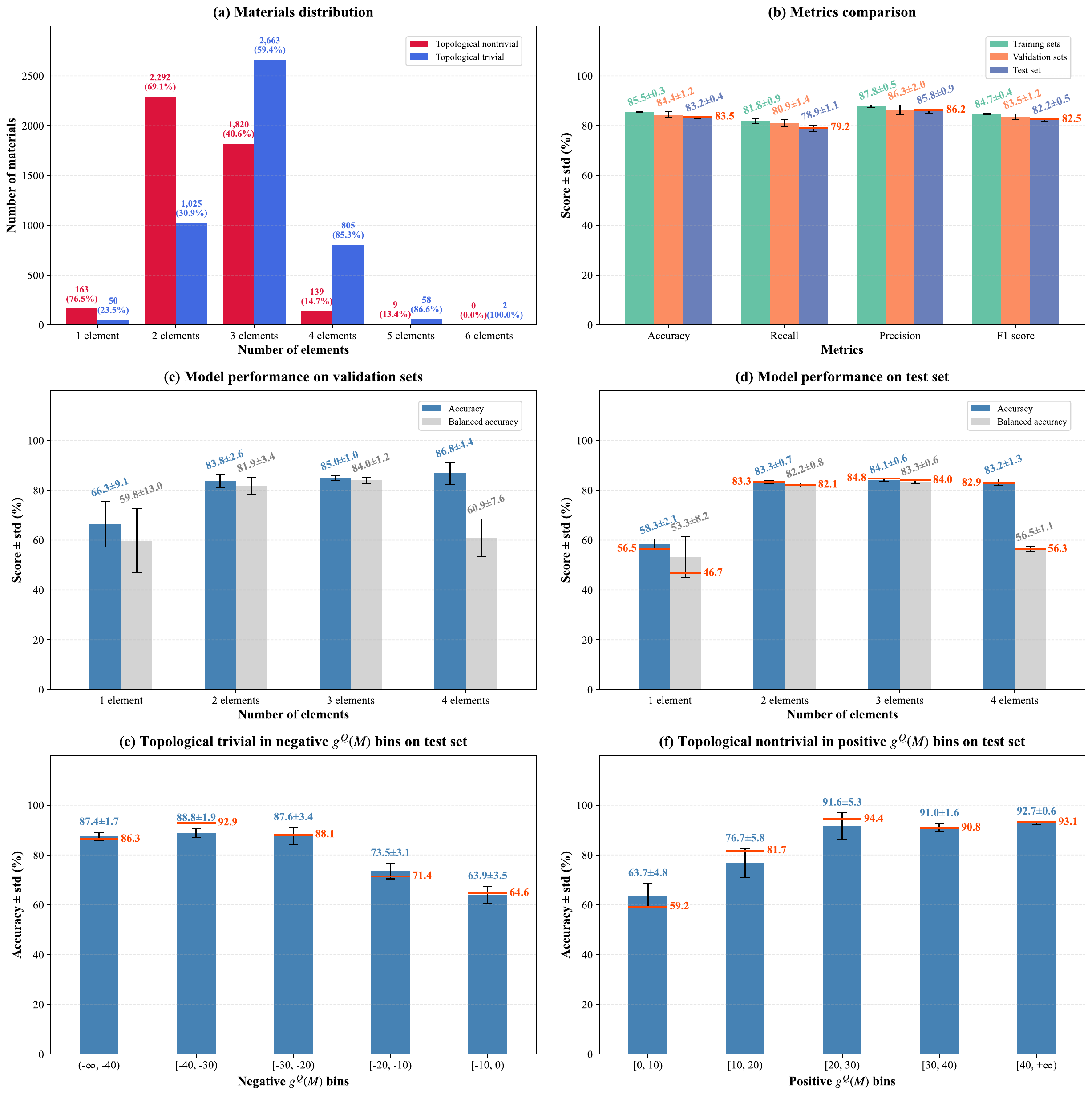}
    \caption{Model performance and dataset analysis. \textbf{(a)} Distribution of labeled materials grouped by elemental complexity (binary, ternary, quaternary, etc.). \textbf{(b)} Summary of model performance in 10-fold cross-validation and on the independent held-out test set (reported as mean $\pm$ standard deviation (std) across folds). \textbf{(c)} Accuracy and balanced accuracy on validation sets, averaged over 10 folds and grouped by material complexity (element count). \textbf{(d)} Accuracy and balanced accuracy on test set, averaged over 10 folds and grouped by material complexity (element count). \textbf{(e)} Test-set accuracy for topologically trivial compounds within negative $g^Q(M)$ bins. \textbf{(f)} Test-set accuracy for topologically nontrivial compounds within positive $g^Q(M)$ bins. In panels (b) and (d)--(f), the red line shows the score obtained on the final test set using the aggregated parameter values employed to compute $g^{Q}(M)$.}
    \label{fig:performance}
\end{figure*}

The dataset contains 9,026 materials, of which 51\% are labeled as topologically trivial and 49\% as nontrivial (combining topological insulators and topological semimetals into a single category), spanning a total of 54 distinct elements. As shown in Figure~\ref{fig:performance}(a), the materials are distributed across compounds containing one to six elements, with counts of 213, 3,317, 4,483, 944, 67, and 2, respectively; the corresponding percentages of topological materials in each group are 76.5\%, 69.1\%, 40.6\%, 14.7\%, 13.4\%, and 0.0\%. As noted in Ref.~\citenum{Ma36}, the dataset is inherently noisy because symmetry-based high-throughput methods cannot capture all topological states, meaning some topological materials may be mislabeled as trivial. In addition, we consider a discovery set of 1,433 materials, also from Ma et al., whose topological character cannot be determined by symmetry indicators. A detailed description of the dataset is given in Section~S3 of the SI. Our approach applies the $g^Q(M)$ chemical rule to this discovery set, followed by DFT validation of materials whose predictions differ from Ref.~\citenum{Ma36}, enabling an assessment of the rule’s predictive performance.

Each dataset entry includes a material’s chemical formula, space group (SG), and topological label, with nontrivial materials labeled 1 and trivial ones labeled 0. Since our chemical rule is symmetry-agnostic, only chemical formulas and labels are used for training. Labels are converted into one-hot vectors—[0,1] for nontrivial and [1,0] for trivial—to facilitate model learning.

\begin{figure*}[tbh]
    \centering
    \includegraphics[width=0.8\linewidth]{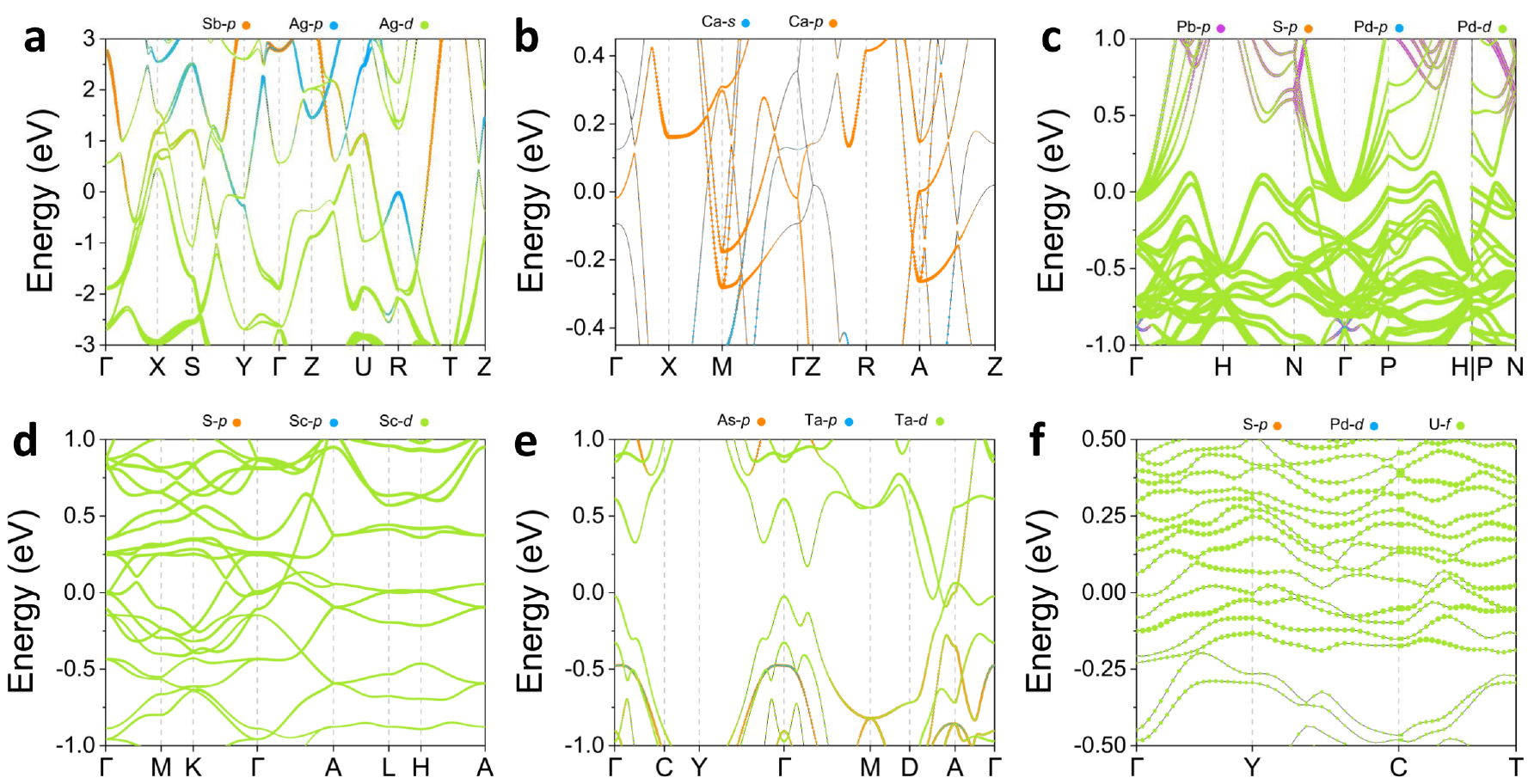}
    \caption{Orbital projected electronic band structures of representative compounds illustrating diverse topological behaviors for (a) Ag$_3$Sb (SG 59), (b) Ca (SG 64), (c) Pd$_3$(PbS)$_2$ (SG 166), (d) P$_3$Sc$_7$ (SG 186), (e) As$_2$Ta (SG 12), and (f) PdS$_4$U$_2$ (SG 43).}
    \label{fig:dft}
\end{figure*}

To evaluate the predictive capability of our quantum-inspired chemical rule, we apply Eq.~\eqref{eq:gm} to classify materials in terms of their topological or trivial character. Prior to screening the 1,433-material discovery set, we assess performance on the full labeled dataset of 9,026 materials. From this dataset, 8,000 materials are randomly selected as the training subset, while the remaining 1,026 compounds serve as an independent hold-out test set.

We train our model using 10-fold cross-validation on the 8,000-material training subset. Each fold produces a full set of learned topogivities and pairwise correlation parameters; these ten parameter sets are subsequently normalized and averaged to obtain the final chemical rule used throughout this work (see Section~S4 of the SI for details).

{\color{black}Figure~\ref{fig:ptable} displays the periodic-table distribution of the learned topogivities. The trends are chemically intuitive: elements with strong spin-orbit coupling—such as Ta, Nb, and Mo—exhibit strongly positive topogivities, consistent with their known tendency to support topological electronic phases. Conversely, light elements (Li, Be, B) contribute only weakly, while nonmetals and halogens show pronounced negative values, reflecting their frequent association with electronically trivial systems. Additional trends are also discernible,  within Group 15, topogivities increase from negative for N, P, As to positive for Bi, reflecting the growing role of spin-orbit coupling. Early transition metals (Groups 4–6) generally show positive values, consistent with their frequent appearance in topological semimetals, while late transition metals exhibit smaller values, likely due to their filled or nearly filled d orbitals and lower tendency to form band inversions.}

Figure~\ref{fig:performance}(b) summarizes the model’s performance across cross-validation and on the final independent test set. The averaged metrics demonstrate consistently strong predictive capacity across folds. On the final 1,026-material test set, the learned chemical rule achieves 83.5\% accuracy, 79.2\% recall, 86.2\% precision, and an F1 score of 82.5\%, indicating reliable generalization to unseen materials. Detailed definitions of these metrics, including the calculation of true positives (TP), true negatives (TN), false positives (FP), and false negatives (FN), are provided in Section~S4 of the SI.

We further analyze accuracy and balanced accuracy as functions of compositional complexity (Figure~\ref{fig:performance}(c)-(d)). Because five- and six-element materials are extremely rare in the labeled dataset, we restrict our analysis to compounds with one to four distinct elements to ensure that each validation fold contains both trivial and topological samples. The model performs most robustly for binary and ternary materials, which constitute the majority of the dataset. For one-element and four-element compounds, the accuracy remains relatively high; however, the balanced accuracy drops significantly for four-element systems. This discrepancy arises from the strong label imbalance evident in Figure~\ref{fig:performance}(a), where the fraction of topological materials decreases markedly as the number of constituent elements increases. As a consequence, accuracy alone becomes misleading, reinforcing the necessity of reporting balanced accuracy in this setting. On the independent test set (Figure~\ref{fig:performance}(d)), the learned rule achieves 83.3\% accuracy for binary materials and 84.8\% for ternary materials. To compare the performance of HQCNN against CVNN and ANN, a detailed analysis is presented in Section~S5 of the SI.

Finally, Figure~\ref{fig:performance}(e)-(f) provides an uncertainty-resolved analysis based on the magnitude of $g^{Q}(M)$. The model exhibits high accuracy and minimal standard deviation for $|g^{Q}(M)| \ge 20$, where classifications are consistently reliable. In the intermediate region $-20 < g^{Q}(M) < 20$, the accuracy decreases and the standard deviation increases, reflecting intrinsically higher classification uncertainty near the decision boundary. This behavior supports the use of $|g^{Q}(M)| \ge 20$ as a practical high-confidence threshold for screening applications. A detailed confidence-score analysis from the topological-phase perspective is provided in Section~S6 of the SI. In particular, for topological predictions ($g^{Q}(M) \ge 20$), the classification accuracy reaches 92.8\% on the independent test set——substantially higher than the overall test set accuracy of 83.5\%——further validating the reliability of this threshold.

For the high-throughput screening stage, we applied our quantum-inspired diagnostic function, $g^{Q}(M)$, to all 1,433 materials in the discovery space. Each material’s topological character was first assessed based on its $g^{Q}(M)$ value. Based on the diagnostic results, among the 142 materials initially identified as topological (i.e., $g^Q(M) > 0$), we screened 79 materials that satisfied the high-confidence topologically nontrivial classification criterion (i.e., $g^Q(M) \ge 20$).

Following the screening strategy adopted in Ref.~\citenum{Ma36}, we excluded seven compounds from the candidate list, resulting in a refined set of 72 high-priority materials for subsequent first-principles validation.

Of these, 66 materials had already been examined in Ref.~\citenum{Ma36}, where 56 were confirmed to be topological. We therefore concentrated our computational efforts on the six previously unvalidated candidates: Ca (SG 64), As$_2$Ta (SG 12), Ag$_3$Sb (SG 59), P$_3$Sc$_7$ (SG 186), Pd$_3$(PbS)$_2$(SG 166), and  PdS$_4$U$_2$ (SG 43). For these compounds, DFT calculations were performed using the VASP package.\cite{kresse1993ab,kresse1996efficiency} Further computational details are provided in the Experiment Section. 

{\color{black}As shown in Figure~\ref{fig:dft}, our calculations verified that five of these six compounds exhibit topological characteristics (Figure~\ref{fig:dft}(a)-(e)), with PdS$_4$U$_2$ identified as a topologically trivial metal (Figure~\ref{fig:dft}(f)). It should be noted that quantitative band-gap values may depend on the choice of exchange-correlation functional. Nevertheless, the topological classifications presented here are supported by first-principles calculations and symmetry indicators from the open-access Topological Materials Database.\cite{Vergniory28,topomaterials} According to the Topological Materials Database, Ca (SG 64) is associated with the symmetry indices $Z_{2w,1} = Z_{2w,2} = Z_{2w,3} = 0$ and $Z_4 = 3$, while both As$_2$Ta (SG 12) and Ag$_3$Sb (SG 59) are listed with indices $Z_{2w,1} = Z_{2w,2} = Z_{2w,3} = 1$ and $Z_4 = 2$. Furthermore, Pd$_3$(PbS)$_2$ (SG 166) is classified in the database as a semimetal with a line crossing type. No entry was found for P$_3$Sc$_7$ (SG 186) in the database, highlighting the relevance of our first-principles calculations in identifying its topological characteristics.}

Altogether, our quantum-inspired diagnostic framework successfully identified 61 topological materials: the 56 previously validated in Ref.~\citenum{Ma36} and five newly discovered through our approach. Furthermore, our method reduces the number of false positives by two relative to the screening workflow of Ref.~\citenum{Ma36}, achieving an overall prediction accuracy of 84.7\%. We view this improvement as a physically motivated extension of the classical topogivity framework, incorporating inter-element correlations to advance beyond the original approach. At the same time, we recognize that the higher accuracy may result from multiple factors, including differences in model architecture and training methodology. It is not that the original method has inherent flaws, but rather that extending the classical framework—motivated by quantum mechanical considerations—to include inter-element correlations can yield meaningful predictive gains.

\section{Experiment}
{\color{black}All computational work, including data processing, model training, and subsequent analysis, was performed within the Python ecosystem. The hybrid quantum–classical  neural network (HQCNN) was implemented using the VQNet framework.\cite{chen2019vqnet, bian2023vqnet} The labeled dataset of 9,026 materials and the discovery set of 1,433 materials were taken from Ma et al.,\cite{Ma36} with the original labels derived from symmetry-indicator-based high-throughput screening of Tang et al.\cite{Tang26} The dataset contains 54 distinct elements. The chemical formula of each material was converted into an elemental composition fraction vector as the input feature for the model. A 10-fold cross-validation was performed on a randomly selected training subset of 8,000 materials, with the remaining 1,026 materials held out as an independent test set. The HQCNN model was trained using the Adam optimizer with a learning rate of 0.01 and a batch size of 64 for 100 epochs. Detailed derivations of  the HQCNN model architecture, the quantum-inspired chemical rule, the complex-valued neural network (CVNN) validation, dataset preprocessing, training protocols, and additional performance analyses are provided in the SI.

First-principles density functional theory (DFT) calculations were performed using the Vienna Ab initio Simulation Package (VASP)\cite{kresse1993ab, kresse1996efficiency} to validate the predicted topological candidates. Core and valence electrons were treated using the projector-augmented wave(PAW) method with a plane-wave cutoff energy of 600 eV.\cite{kresse1999from} Structural relaxations were converged until forces were below 0.001 eV/{\AA} and total-energy changes were less than $10^{-8}$ eV. A k-mesh commensurate with each lattice was adopted. Spin-orbit coupling (SOC) was included in all electronic structure calculations. 
}

\section{Concluding Remarks}

In this work, we have introduced a quantum-inspired chemical rule derived from a hybrid quantum–classical neural network, extending the classical machine-learning-based topogivity framework into the quantum domain. In this formulation, the model structure reflects how information is processed in quantum systems, and the emergence of pairwise correlation terms arises naturally from quantum interference effects—features that have no direct counterpart in conventional classical approaches.

This perspective is particularly relevant in the current era of rapidly advancing quantum computing, where a central goal is to reformulate classical scientific models into quantum-compatible forms. Our results demonstrate how a classical compositional rule transforms when embedded within a quantum framework, revealing quantum-specific interaction terms that enhance the representation of inter-element correlations.

From a materials discovery standpoint, the proposed rule serves as an efficient high-throughput screening tool that requires only compositional information while enabling accurate and scalable prediction of topological properties. By pre-selecting high-confidence candidates, it significantly reduces the computational and experimental effort required for materials exploration, including systems beyond the scope of symmetry-indicator methods.

Looking forward, such quantum-inspired formulations provide a foundation for future implementation on quantum hardware, offering new opportunities to explore complex correlations in materials systems. More broadly, this work highlights a promising pathway for integrating quantum computing with materials informatics, accelerating the discovery and design of functional quantum materials.

Future efforts may extend this framework in several directions. First, integrating additional descriptors—such as space group, total and valence electron counts, or chemical family information (e.g., alkali metals, halogens)—could further improve classification accuracy. Incorporating such features would not only enhance predictive performance but also mitigate compositional leakage by providing richer representations that distinguish materials with identical stoichiometry but different structures or chemical contexts. This would naturally account for family-level similarities between training and test materials, reducing the risk of overly optimistic generalization estimates.

Second, applying the quantum-inspired rule to larger and more diverse materials databases—particularly those with reduced label noise—will expand its discovery potential while deepening our understanding of elemental contributions and quantum-originated cross-correlation patterns in topological materials. Although the mathematical form and physical motivation of the pairwise interference terms are clear, their specific quantitative contributions and underlying physicochemical interpretations remain subjects for further investigation. A deeper understanding of when and why these interference terms become significant—and what specific physical mechanisms they represent—will require more extensive analysis as the model is applied to broader classes of materials. We view this as an important direction for future research, bridging the gap between quantum-inspired modeling and physical interpretability.

\section{Supporting Information}
Derivation of quantum-inspired chemical rules based on the HQCNN model, validation with CVNN, dataset details,  training protocols, performance comparisons across models, and confidence analysis of the $g^Q(M)$ score.

\section{Code Availability}
The code used to obtain the described results, will be openly released upon acceptance of the manuscript at  \url{https://github.com/Arif-PhyChem/quantum_inspired_chemical_rule}

\section{Acknowledgments}
A.U. acknowledges funding from the National Natural Science Foundation of China (No. W2433037) and the Natural Science Foundation of Anhui Province (No. 2408085QA002). 
M. Y. acknowledges funding from Scientific Research Projects of Anhui Provincial Department of Education under Grant Nos. 2025AHGXZK20126 and 2025AHGXZK50054. P.O.D. acknowledges funding by the projects for International Senior Scientists (Project No.: W2531013) and for Outstanding Youth Scholars (Overseas, 2021) of the National Natural Science Foundation of China, via the Lab project of the State Key Laboratory of Physical Chemistry of Solid Surfaces.

\bibliography{AMI_revise}

\end{document}


\maketitle

\vspace{2em}
\begin{center}
\renewcommand{\contentsname}{Table of Contents}
\tableofcontents
\end{center}
\vspace{1em}
\hrule
\vspace{1em}

\section{Derivation of Quantum-Inspired Chemical Rule from Hybrid Quantum–Classical Neural Network}
\begin{figure}[tbh]
    \centering
    \includegraphics[width=\linewidth]{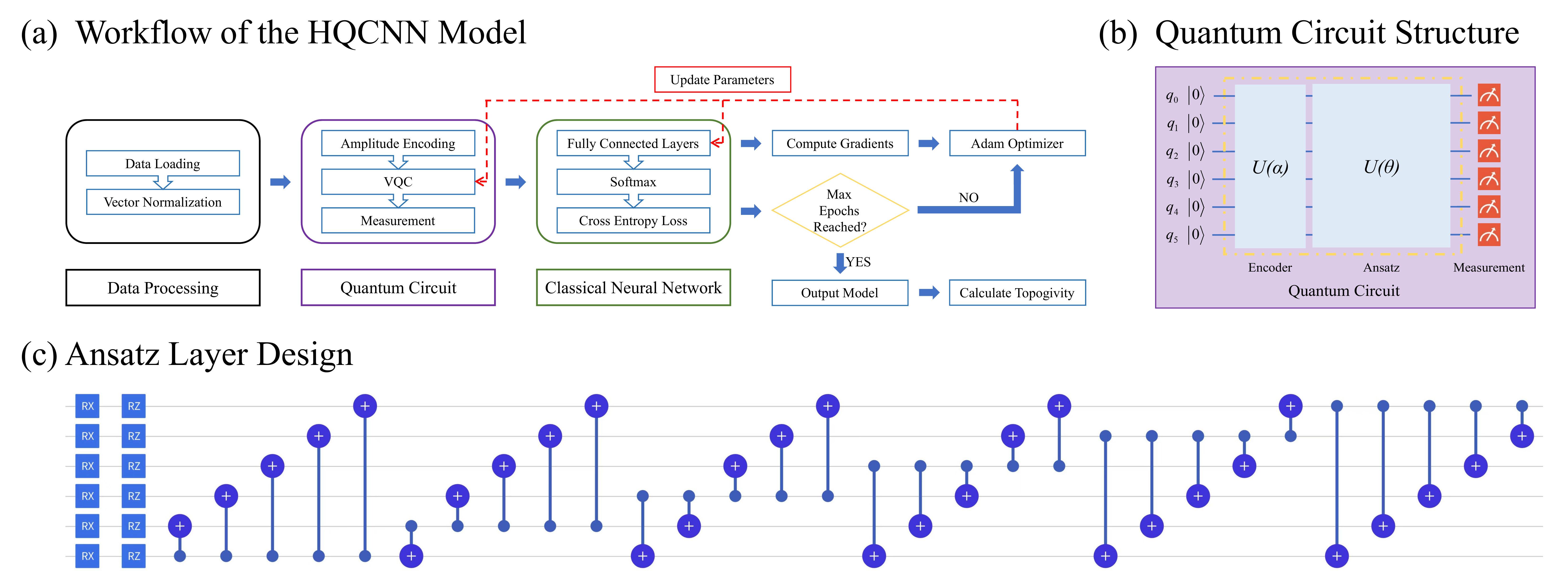}
    \caption{Schematic of the HQCNN model and quantum circuit. \textbf{(a)} In our HQCNN model, the workflow begins with vector normalization of the loaded data, which is then encoded into a parameterized VQC using amplitude encoding. The output state of VQC is measured, and the results are fed into a classical fully connected neural network. The network's output is activated by the softmax function, and the cross-entropy loss is computed. The Adam optimizer minimizes this loss via backpropagation, simultaneously optimizing the parameters in both the VQC and the classical network. After a fixed number of epochs (epoch=100), the final model and parameters are saved. These parameters are subsequently used to calculate the topogivity. \textbf{(b)} The quantum circuit comprises six qubits ($q_{0} - q_{5}$). The encoder employs the amplitude encoding method, with its circuit corresponding to the unitary matrix $U(a)$, while the ansatz forms the parameterized part of the circuit, corresponding to the unitary matrix $U(\theta)$. Finally, the probability distribution is obtained by measuring the six qubits. \textbf{(c)} The depicted layer shows the fundamental structure of the Ansatz circuit. Each horizontal line in the layer represents a qubit. The complete Ansatz comprises four identical such layers, each consisting of parameterized Rx and Rz gates, along with CNOT gates that act as entangling operations.}
    \label{fig:HQCNN}
\end{figure}

In this section, we outline how the chemical rule (Eq.~1 of our main text) emerges from our proposed hybrid quantum–classical neural network (HQCNN). Our HQCNN consists of two tightly integrated components—a quantum module and a classical module—as illustrated in Figure~\ref{fig:HQCNN}.

The quantum module begins with the normalization of input features, which are then encoded into a parameterized Variational Quantum Circuit (VQC) using amplitude encoding. The circuit comprises two main parts: an Encoder, responsible for embedding the classical data into quantum states, and an Ansatz, which forms the trainable section of the circuit. The Ansatz is composed of four identical layers, each containing parameterized Rx and Rz rotation gates, along with CNOT gates that introduce entanglement between qubits. The qubits are subsequently measured, and the measurement outcomes form the latent quantum representation of the material.

These measured outputs are passed to the classical module, a fully connected neural network that processes the quantum features to predict material properties. The final output layer employs a softmax activation, while model optimization is performed using the cross-entropy loss and Adam optimizer, which jointly update parameters in both the VQC and classical layers through backpropagation. After training for a fixed number of epochs, the optimized parameters are utilized to compute topogivity values for each material.

As input to the HQCNN model, we define for each material $M$ an elemental composition fraction vector, $\vec{F} (M)$, whose components $f_{E}(M)$ represent the fractional abundance of each element $E$ in the compound. Elements absent from a material are assigned zero values, whereas those not present in the dataset are excluded from the vector. For the labeled dataset used in this work, $\vec{F} (M)$ is a 54-dimensional vector, corresponding to all unique elements considered, ordered by ascending atomic number for convenience. Mathematically, it can be expressed as:

\begin{align}\vec{F} (M)=(f_{Li}(M),f_{Be}(M),f_{B}(M),f_{C}(M),\dots ,f_{U}(M))^{T} \, .\end{align}

For notational simplicity, we denote the $j$-th component of $\vec{F} (M)$ as $f_{j}(M)$, such that:

\begin{align}\vec{F} (M)=(f_{0}(M),f_{1}(M),f_{2}(M),f_{3}(M),\dots ,f_{53}(M))^{T} \, . \end{align}

We then encode $\vec{F} (M)$ into the quantum circuit via amplitude encoding, which requires taking the square root of each component $f_{j}(M)$ in vector $\vec{F} (M)$. Given that the vector comprises 54 components, a minimum of 6 qubits is necessary (as 6 qubits can encode $2^{6} =64$ data points). This results in 10 unoccupied positions, necessitating a zero-padding operation on $\vec{F} (M)$ by appending 10 zeros, and denote the resulting 64-dimensional vector as $\vec{f} (M)$ leading to:
\begin{align}\vec{f} (M)=(\textstyle\sqrt{f_{0}(M)},\dots ,\textstyle\sqrt{f_{53}(M)},0,\dots,0)^{T} \, . \end{align}

The normalized vector $\vec{f} (M)$ is encoded into the quantum circuit through amplitude encoding, initializing from the $\left | 000000 \right \rangle$  computational basis state. The resulting quantum state is prepared as:

\begin{equation}\begin{aligned}\left | \psi_{in}   \right \rangle=&\textstyle\sqrt{f_{0}(M) } \left | 000000  \right \rangle+ \dots +\textstyle\sqrt{f_{53}(M) } \left | 110101  \right \rangle \\&+0\cdot \left | 110110  \right \rangle+\dots + 0\cdot \left | 111111   \right \rangle \, ,
\end{aligned}
\end{equation}
where $\sum_{j} \left| \textstyle\sqrt{f_{j}(M)} \right|^2=1$. As shown in Figure~\ref{fig:HQCNN}, the parameterized Ansatz of our HQCNN employs a fully connected quantum circuit, chosen for its balance between high expressibility and low parameter count. The optimized Ansatz acts as a unitary transformation on the input quantum state $\left | \psi_{in}   \right \rangle$, such that:

\begin{align}U(\theta) \left | \psi_{in}   \right \rangle= \left | \psi_{out}  \right \rangle \, , 
\end{align}
which can be expanded into the following form:

\begin{equation} \label{eq:mat}
\left[\begin{array}{ccc}
u_{0,0}+v_{0,0}i  &  \dots & u_{0,63}+v_{0,63}i\\
\vdots & \ddots  & \vdots  \\
u_{63,0}+v_{63,0}i  & \dots & u_{63,63}+v_{63,63}i\\
\end{array}\right]\begin{bmatrix}
\textstyle\sqrt{f_{0}(M) }  \\
\vdots \\
\textstyle\sqrt{f_{53}(M) } \\
\hline
0 \\
\vdots \\
0
\end{bmatrix} \, ,
\end{equation}
further leading to the output quantum state:

\begin{equation}\begin{aligned} \left | \psi_{out}   \right \rangle=C_{0}\left |000000  \right \rangle+C_{1}\left |000001  \right \rangle+\dots +C_{63}\left |111111  \right \rangle \, , 
\end{aligned}\end{equation}
where each complex amplitude is given by:

\begin{equation}
\begin{aligned}
C_{n}&=\sum_{j=0}^{53} (u_{n,j}+v_{n,j}i) \textstyle\sqrt{f_{j}(M)}\\
&= \sum_{j=0}^{53} u_{n,j} \textstyle\sqrt{f_{j}(M)}\displaystyle + \sum_{j=0}^{53}v_{n,j}  \textstyle\sqrt{f_{j}(M)}i\\
&= u_{n}+v_{n}i \, .
\end{aligned}
\end{equation}

Upon measurement, the quantum circuit produces an output probability vector $\vec{P}$ with 64 components, where each component is given by:

\begin{equation}
\begin{aligned}
\left | C_{n}  \right | ^{2}=& \left | u_{n}  \right | ^{2}+\left | v_{n}  \right | ^{2}\\
=&\sum_{j=0}^{53}(u_{n,j}^{2} +v_{n,j}^{2} )f_{j} (M)\\
&+ 2 \sum_{j=0}^{52} \sum_{k=j+1}^{53} (u_{n,j}u_{n,k}+v_{n,j}v_{n,k} )\textstyle\sqrt{f_{j}(M)f_{k}(M) } \, .
\end{aligned}
\end{equation}
where $n=0,1,2,\dots,63$. Please note that the second summation term introduces cross terms reflecting interactions among different features—an intrinsic effect arising from quantum measurement. Finally, the resulting probability vector $\vec{P}$ is then used as input to the subsequent classical artificial neural network (ANN) module.

For the classical ANN module of our HQCNN model, we adopt a three-layer fully connected architecture. The input layer comprises 64 neurons, corresponding to the 64 components $\left| C_n \right|^2(n=0,1,2,\dots,63)$ of the probability vector $\vec{P}$ obtained from the quantum module. The hidden layer contains 24 neurons, while the output layer consists of 2 neurons. A softmax activation function is applied at the output stage to produce normalized classification probabilities. While activation functions are applied during training, they are not strictly necessary for inference in routine material topology prediction. Importantly, the ANN employs only weight parameters ($w$) and omits bias terms ($b$), simplifying the network and preserving interpretability.

Data transformation in the ANN occurs through hierarchical propagation across layers, which can be compactly represented as a sequence of matrix multiplications. Let $W_1$ denote the weight matrix connecting the input and hidden layers, and $W_2$ the weight matrix connecting the hidden and output layers. Then, for an input vector $\vec{C}$, the network output can be expressed as:

\begin{equation}
    \vec{y} = W_2 (W_1 \vec{C}) = (W_2 W_1) \vec{C} \, ,
\end{equation}
demonstrating that the overall transformation of the network can be represented by a single effective weight matrix:

\begin{equation}
\begin{aligned}
W_{\text{ANN}} & = W_2 W_1 \\
&= \begin{bmatrix}
  u_{2,0,0} & \dots & u_{2,0,23} \\
  u_{2,1,0} & \dots & u_{2,1,23} 
\end{bmatrix} 
\begin{bmatrix}
 u_{1,0,0} & \dots & u_{1,0,63} \\
  \vdots & \ddots  & \vdots \\
  u_{1,23,0} & \dots & u_{1,23,63}
\end{bmatrix}  \\
&=\begin{bmatrix}
  w_{0,0} & \dots & w_{0,63}\\
  w_{1,0} & \dots & w_{1,63}
\end{bmatrix} \, .
\end{aligned}
\end{equation}

We apply this overall parameter matrix $W_{\text{ANN}}$ to the vector $\vec{P}$, obtaining:
\begin{equation}
\begin{aligned}W_{ANN} \vec{P} & = \begin{bmatrix}  w_{0,0} &\dots   & w_{0,63}\\  w_{1,0} &\dots   & w_{1,63}\end{bmatrix}  \begin{bmatrix} \left | C_{0}  \right |^{2}  \\ \vdots \\ \left | C_{63}  \right |^{2} \end{bmatrix}\\&=\begin{bmatrix} h_{0} \\ h_{1}\end{bmatrix} \, , \end{aligned}
\end{equation}
where $h_{0}=\sum_{n=0}^{63} w_{0,n}\left | C_{n}  \right |^{2}$ , $h_{1}=\sum_{n=0}^{63} w_{1,n}\left | C_{n}  \right |^{2}$. These two outputs correspond to the two classification categories of the network (trivial vs. topological). To define a single diagnostic score for material $M$, we take the difference between these outputs:

\begin{equation} \label{eq:final_result}
\begin{aligned}
h(M)  =& h_{1}-h_{0} \\
 =& \sum_{n=0}^{63}(w_{1,n}-w_{0,n})[\sum_{j=0}^{53}(u_{n,j}^{2} +v_{n,j}^{2} )f_{j} (M) \\
&+ 2 \sum_{j=0}^{52} \sum_{k=j+1}^{53} (u_{n,j}u_{n,k}+v_{n,j}v_{n,k} )\textstyle\sqrt{f_{j}(M)f_{k}(M) }] \\
 =& {\sum_{E} {f_{E}(M)}\tau _{E}}+{\sum_{E}\!\sum_{\substack{E' \\ E' \neq E}}\! {\textstyle\sqrt{f_{E}(M)f_{E^{'} }(M) } }\tau _{EE^{'} }} \\
 =& g(M)+\!{\sum_{E\ne   E^{'} }\! {\textstyle\sqrt{f_{E}(M)f_{E^{'} }(M) } }\tau _{EE^{'} }} \\
 =& g^Q(M) \, , 
\end{aligned}
\end{equation}
{\color{black}where the first sum runs over all elements appearing in the chemical formula of material $M$, and the double sum runs over all distinct pairs of elements, with $E$ and $E^{'}$ ordered such that the atomic number of $E$ is less than that of $E^{'}$. $f_E(M)$ and $f_{E^{'}}(M)$ denote the fractional abundances of elements $E$ and $E^{'}$, respectively. The coefficients $\tau_E$ and $\tau_{EE^{'}}$ are given by:}

\begin{align}
       \tau _{E} & =\sum_{n=0}^{63}(w_{1,n}-w_{0,n})(u_{n,E}^{2} +v_{n,E^{'}}^{2} )\, , \\
       \tau _{EE^{'} } & =\sum_{n=0}^{63}2(w_{1,n}-w_{0,n})(u_{n,E}u_{n,E^{'}}+v_{n,E}v_{n,E^{'}} ) \, .
\end{align}
Here, the indices $E$ and $E^{'}$ (with $E^{'}$ running over all elements distinct from $E$) denote distinct chemical elements within a compound, replacing the earlier indices $j$ and $k$. The term $\tau _{E}$ describes the self-contribution of element $E$, while $\tau _{EE^{'} }$ quantifies pairwise correlations between different elements $E$ and $E^{'}$. 

Equation~\eqref{eq:final_result} thus yields the quantum-inspired chemical rule $g^Q(M)$ of Eq.~1 (main text). Notably, the same rule—including the correlation terms—can be derived using an equivalent CVNN model, which reproduces the exact analytical structure obtained from the HQCNN. For completeness, the full derivation is provided in the next section.

\section{Validation of the Quantum-Inspired Chemical Rule Using Complex-Valued Neural Networks}

This appendix aims to provide an independent theoretical verification for the quantum-inspired chemical rule extended in the main text by constructing a complex-valued neural network (CVNN). In the HQCNN framework developed in this work, the core of the quantum circuit is composed of complex unitary matrices, whose weights are inherently complex, naturally enabling the expression of quantum phase interference and coherent superposition. This physical background suggests that if a model is constructed in a purely data-driven manner, complex weights likewise serve as an ideal mathematical mechanism for capturing the deep-seated correlations between material composition and topological properties.

The essence of the CVNN constructed here lies in extending the weights to the complex domain. This is consistent with the mathematical nature of quantum circuits, as both encode phase information and rotational transformations beyond real space through complex operations. This structural homology strongly implies that both the physics-based quantum model and the data-driven complex network lead to the same conclusion: the chemical rules governing the topological properties of materials inherently involve coupling relationships at the complex level among elemental interactions. Therefore, the emergence of the pairwise interaction terms in the chemical rule of Eq.~1 is not accidental but instead a natural consequence of the underlying complex representation shared by both models. To substantiate this claim, we explicitly reconstruct Eq.~1 using the CVNN formalism below.

We begin by specifying the input to the CVNN. For each material $M$, we introduce an elemental composition fraction vector, $\vec{F} (M)$, whose components $f_{E}(M)$ represent the fractional abundance of each element $E$ in the compound, just as we did in the HQCNN model. Elements absent from a material are assigned zero values, whereas those not present in the dataset are excluded from the vector. For the labeled dataset used in this work, $\vec{F} (M)$ is a 54-dimensional vector, corresponding to all unique elements considered, ordered by ascending atomic number for convenience. Mathematically, it can be expressed as:

\begin{align}\vec{F} (M)=(f_{Li}(M),f_{Be}(M),f_{B}(M),f_{C}(M),\dots ,f_{U}(M))^{T} \, .\end{align}

For notational simplicity, we denote the $j$-th component of $\vec{F} (M)$ as $f_{j}(M)$, such that:

\begin{align}\vec{F} (M)=(f_{0}(M),f_{1}(M),f_{2}(M),f_{3}(M),\dots ,f_{53}(M))^{T} \, . \end{align}

Unlike the HQCNN, the input to the CVNN is not constrained by the dimensionality imposed by the number of qubits, so we do not need to zero-pad $\vec{F} (M)$. Furthermore, since the input to the CVNN must be in complex form (a+bi), we convert each component $f_{j}(M)$ in $\vec{F} (M)$ into the complex form $\textstyle\sqrt{\frac{f_{j}(M)}{2} } + \textstyle\sqrt{\frac{f_{j}(M)}{2}} i$. Consequently, the input vector $\vec{f} (M)$ takes the following form:

\begin{align}
\vec{f}(M) & = (\textstyle\sqrt{\frac{f_{0}(M)}{2} } + \textstyle\sqrt{\frac{f_{0}(M)}{2}}i \, ,
\dots ,\textstyle\sqrt{\frac{f_{53}(M)}{2} } + \textstyle\sqrt{\frac{f_{53}(M)}{2}}i)^{T} \, ,
\end{align}
where$\sum_{j} \left | \textstyle\sqrt{\frac{f_{j}(M)}{2} } + \textstyle\sqrt{\frac{f_{j}(M)}{2}}i \right |^2 =1 $ .

Our CVNN model adopts a three-layer complex-valued fully-connected architecture. The input layer contains 54 neurons, the hidden layer has 15 neurons, and the output layer consists of 2 neurons. The network outputs a pair of complex numbers. We calculate their squared moduli, which are then processed by a softmax activation function to generate normalized classification probabilities. Although the softmax activation function is used during training, it is not strictly necessary for inference in conventional material topology prediction. It is worth noting that this complex-valued neural network uses only weight parameters (w) and omits bias terms (b), thereby simplifying the network structure and maintaining interpretability. 

Similar to any artificial neural network (ANN), data transformation in the CVNN is achieved through layer-wise propagation, which can be compactly represented as a sequence of matrix multiplications. Let W1 denote the complex-valued weight matrix connecting the input and hidden layers, and W2 denote the complex-valued weight matrix connecting the hidden and output layers. Then, for an input vector C, the network output can be expressed as:

\begin{align}
    \vec{y} = W_2 (W_1 \vec{C}) = (W_2 W_1) \vec{C} \, ,
\end{align}

demonstrating that the overall transformation of the network can be represented by a single effective weight matrix:

\begin{align}
W_{\rm CVNN} & = W_{2}W_{1} \nl
&=\begin{bmatrix}
 u_{2,0,0}+v_{2,0,0}i  &  \dots  & u_{2,0,14}+v_{2,0,14}i\\
 u_{2,1,0}+v_{2,1,0}i  &  \dots  & u_{2,1,14}+v_{2,1,14}i
\end{bmatrix} \nl
& \;\;\;\;\;\times 
 \begin{bmatrix}
 u_{1,0,0}+v_{1,0,0}i & \dots & u_{1,0,53}+v_{1,0,53}i\\
 \vdots  & \ddots  & \vdots \\
 u_{1,14,0}+v_{1,14,0}i & \dots & u_{1,14,53}+v_{1,14,53}i
\end{bmatrix}\nl 
&=\begin{bmatrix}
 u_{0,0}+v_{0,0}i & \dots & u_{0,53}+v_{0,53}i\\
 u_{1,0}+v_{1,0}i & \dots & u_{1,53}+v_{1,53}i
\end{bmatrix}\, ,
\end{align}
We apply this overall parameter matrix $W_{\text{CVNN}}$ to the vector $\vec{f}(M)$, obtaining:

\begin{align}
W_{\rm CVNN}\vec{f}(M) & = \begin{bmatrix}
 u_{0,0}+v_{0,0}i & \cdots & u_{0,53}+v_{0,53}i\\
 u_{1,0}+v_{1,0}i & \cdots & u_{1,53}+v_{1,53}i
\end{bmatrix}  \nl 
&\;\;\;\;\; \times 
\begin{bmatrix}
\sqrt{ \frac{f_{0}(M)}{2} } + \sqrt{\frac{f_{0}(M)}{2} }i \\
\vdots\\
\sqrt{\textstyle\frac{f_{53}(M)}{2} } + \sqrt{\frac{f_{53}(M)}{2} }i
\end{bmatrix}\nl
&=\begin{bmatrix}
\frac{1}{\sqrt{2} }(\sum\limits_{j=0}^{53}(u_{0,j}-v_{0,j})\textstyle \sqrt{f_{j}(M)} + \sum\limits_{j=0}^{53}(u_{0,j}+v_{0,j}) \sqrt{f_{j}(M)}i)\\
\frac{1}{\sqrt{2} }(\sum\limits_{j=0}^{53}(u_{1,j}-v_{1,j})\textstyle \sqrt{f_{j}(M)} + \sum\limits_{j=0}^{53}(u_{1,j}+v_{1,j}) \sqrt{f_{j}(M)}i)
\end{bmatrix}\nl
&=\begin{bmatrix}
 h_{0} \\
 h_{1}
\end{bmatrix}\, .
\end{align}

Then, by taking the squared modulus of $h_{0}$ and $h_{1}$, we obtain:
\begin{equation}
\begin{aligned}
\left | h_{n}  \right | ^{2} & = \left | h_{n.real}  \right | ^{2} + \left | h_{n.imag}  \right | ^{2}\\
&=\sum_{j=0}^{53} (u_{n,j}^{2} + v_{n,j}^{2} )f_{j}(M) \\
&\quad +2\sum_{j=0}^{52} \sum_{k=j+1}^{53}(u_{n,j}u_{n,k} + v_{n,j}v_{n,k})\textstyle\sqrt{f_{j}(M)f_{k}(M) } \, .
\end{aligned}
\end{equation}

These two outputs correspond to the two classification categories of the network (trivial vs. topological). To define a single diagnostic score for material $M$, we take the difference between these outputs:

\begin{align}
h(M)  = &\left | h_{1}  \right | ^{2} -  \left | h_{0}  \right | ^{2}\nl
=&\sum_{j=0}^{53} (u_{1,j}^{2} + v_{1,j}^{2} -u_{0,j}^{2} - v_{0,j}^{2})f_{j}(M)\nl
&+2\sum_{j=0}^{52} \sum_{k=j+1}^{53}(u_{1,j}u_{1,k}+v_{1,j}v_{1,k} \nl
&\qquad \qquad -u_{0,j}u_{0,k}-v_{0,j}v_{0,k})\textstyle\sqrt{f_{j}(M)f_{k}(M)}\nl
=& {\sum_{E}f_{E}(M)\tau _{E}}+{\sum_{E}\!\sum_{\substack{E' \\ E' \neq E}} \!{\textstyle\sqrt{f_{E}(M)f_{E^{'} }(M) } }\tau _{EE^{'} }}\nl
 =& g(M)+\!{\sum_{E\ne   E^{'} }\! {\textstyle\sqrt{f_{E}(M)f_{E^{'} }(M) } }\tau _{EE^{'} }}\nl
 =& g ^{Q}(M)\, ,
\end{align}

where the first sum runs over all elements appearing in the chemical formula of material $M$, and the double sum runs over all distinct pairs of elements, with $E$ and $E^{'}$ ordered such that the atomic number of $E$ is less than that of $E^{'}$. $f_E(M)$ and $f_{E^{'}}(M)$ denote the fractional abundances of elements $E$ and $E^{'}$, respectively. The coefficients $\tau_E$ and $\tau_{EE^{'}}$ are given by:

\begin{align}
\tau _{E} & = u_{1,E}^{2} + v_{1,E}^{2} - u_{0,E}^{2} - v_{0,E}^{2}
\,,\\
\tau _{EE^{'}} & = u_{1,E}u_{1,E^{'}} + v_{1,E}v_{1,E^{'}} - u_{0,E}u_{0,E^{'}} - v_{0,E}v_{0,E^{'}}\,. 
\end{align}

Here, the indices $E$ and $E^{'}$ (with $E^{'}$ running over all elements distinct from $E$) denote distinct chemical elements within a compound, replacing the earlier indices $j$ and $k$. The term $\tau _{E}$ describes the self-contribution of element $E$, while $\tau _{EE^{'} }$ quantifies pairwise correlations between different elements $E$ and $E^{'}$. 

The above derivation leads directly to the same chemical rule obtained in Eq.~\eqref{eq:final_result}, providing an independent mathematical confirmation from the complex-valued network. The CVNN not only recovers the same pairwise coupling structure as the HQCNN model but also reveals phase information in the learned complex weights, suggesting that elemental interactions may involve coordinated “synergistic phases” rather than purely additive effects.

Moreover, the complex formulation highlights limitations of real-valued models, which cannot naturally capture such non-commutative interaction patterns. Thus, both the quantum-inspired model and the CVNN converge on the central conclusion: element-pair couplings are indispensable for accurately diagnosing topological materials and form a solid foundation for composition-based topological materials design.

\section{Dataset Description}
The dataset used in this work comes from the labeled dataset compiled by Ma et al.,\cite{Ma36} which itself was curated from the high-throughput symmetry-indicator database of Tang et al.\cite{Tang26} Following the same preprocessing procedure described in Ref.~\cite{Ma36}, we removed materials containing rare elements (defined as those appearing fewer than 25 times in the dataset), leaving 54 distinct elements in total.The chemical formula of each material was converted into an elemental composition fraction vector as the input feature for the model.

The full labeled dataset contains 9,026 materials, with 51\% labeled as trivial and 49\% as topological (here we group topological insulators and topological semimetals together). The dataset covers a wide range of material complexities—binaries, ternaries, and quaternaries—as shown in Figure~3(a) of the main text. We also adopt from Ref.~\cite{Ma36} an independent discovery set of 1,433 materials whose topological nature cannot be determined by symmetry indicators. This set is used for high-throughput screening and subsequent DFT validation.

As discussed in Ref.~\cite{Ma36}, the labels in this dataset should be treated as noisy. The noise comes from several sources.

First, the symmetry-indicator-based high-throughput methods used to generate the original database \cite{Tang26} are inherently limited—they cannot detect certain types of topological phases. As a result, some truly topological materials may be mislabeled as trivial, and vice versa.

Second—and this is particularly relevant to our composition-only approach—materials with the same chemical formula but different space groups can end up with different topological labels. After preprocessing, each entry in the labeled dataset is uniquely identified by the combination of its reduced formula and space group (see Ref.~\cite{Ma36}, Supplementary Section S3.A). However, because our model takes only elemental composition as input and discards all spatial information, it has no way of distinguishing between two materials that share the same reduced formula but belong to different space groups and carry different labels. In such cases, the model is forced to predict the same class for both, inevitably making errors. Ref. [1] quantifies this inherent limitation, showing that at least 203 errors in the labeled dataset are unavoidable due solely to this composition-label ambiguity—corresponding to a maximum achievable accuracy of 97.75\%.

This label noise, together with the inherent ambiguity of using only compositional features, poses real challenges for any supervised learning approach trained on this data. Our model's performance should be understood in light of these limitations. Nevertheless, as shown in the main text, our quantum-inspired rule is still able to extract meaningful chemical trends from this noisy data, and the DFT validation confirms its predictive power on previously unpredicted materials (by the counterpart classical approach).

\section{Training and Evaluation Details}

The HQCNN model presented in this work is implemented using VQNet, Origin Quantum’s hybrid quantum-classical machine learning framework.\cite{chen2019vqnet,bian2023vqnet}{\color{black}The HQCNN model was trained using the Adam optimizer. After preliminary experimentation balancing performance and training efficiency, we selected a learning rate of 0.01 and a batch size of 64. These hyperparameters yielded stable convergence and strong predictive performance across all cross-validation folds for 100 epochs.} As described in the main text, our full labeled dataset contains 9,026 materials. {\color{black}The full dataset was first randomly shuffled. From this shuffled dataset, we randomly selected 8,000 materials as the training subset and set aside the remaining 1,026 as an independent held-out test set. To obtain a reliable estimate of model performance and to mitigate overfitting, we applied 10-fold cross-validation to the 8,000-material training subset. Specifically, we divided the 8,000 samples into 10 equal folds (800 materials each). For each of the 10 cross-validation rounds, one fold was designated as the validation set, while the remaining nine folds (7,200 materials) served as the training set}. Importantly, in every fold the model was reinitialized with a new random seed and trained entirely from scratch. This procedure produced ten independent sets of learned parameters and ten corresponding validation-set performance metrics.

The validation-set scores reported in our work are the average and standard deviation computed over these ten folds. Thus, each validation score represents a fold-by-fold statistic reflecting how well the model generalizes when trained on 90\% of the training data and evaluated on the remaining 10\%.

The detailed performance of the model is presented in Figure~3 of the main text. Here, we define the metrics used throughout this work, all of which are based on the following convention: the positive label corresponds to topological materials ($y = +1$), and the negative label corresponds to trivial materials ($y = -1$). For a given material $M$, let $\hat{y}(M)$ denote the model's predicted label. We define:
\begin{itemize}
    \item \textbf{True Positives (TP):} number of samples with $\hat{y}(M) = +1$ and $y = +1$
    \item \textbf{False Positives (FP):} number of samples with $\hat{y}(M) = +1$ and $y = -1$
    \item \textbf{True Negatives (TN):} number of samples with $\hat{y}(M) = -1$ and $y = -1$
    \item \textbf{False Negatives (FN):} number of samples with $\hat{y}(M) = -1$ and $y = +1$
\end{itemize}

Based on these, we compute the following metrics:

\begin{equation}
\text{Accuracy} = \frac{TP + TN}{TP + TN + FP + FN}
\end{equation}

\begin{equation}
\text{Precision} = \frac{TP}{TP + FP}
\end{equation}

\begin{equation}
\text{Recall} = \frac{TP}{TP + FN}
\end{equation}

\begin{equation}
F_1 \text{ score} = 2 \times \frac{\text{Precision} \times \text{Recall}}{\text{Precision} + \text{Recall}}
\end{equation}

\begin{equation}
\text{Balanced Accuracy} = \frac{1}{2} \left( \frac{TP}{TP + FN} + \frac{TN}{TN + FP} \right)
\end{equation}

All metrics are reported separately for the validation sets (averaged over 10 folds with standard deviation) and the independent test set.

After completing the cross-validation, we combined the ten independently learned parameter sets into a single, final parameter table. Before averaging, each parameter set was normalized relative to the oxygen topogivity value, $\tau_{\mathrm{O}}$, which is particularly stable because oxygen appears frequently in the dataset. We first computed the mean oxygen topogivity  $\tau_{\mathrm{O,avg}}$ across all ten folds. Then, for each parameter set, we calculated a scaling factor equal to $\tau_{\mathrm{O,avg}}$ divided by its $\tau_{\mathrm{O}}$. This scaling factor was applied to all 1,485 parameters in that fold, aligning all parameter sets to a consistent oxygen-normalized reference. Finally, we performed an element-wise average across the ten normalized sets, yielding the final topogivities and inter-element correlation parameters.

These final averaged parameters—not a separately trained model—are then used to compute $g^{Q}(M)$ for any new material. The final test-set scores reported in the main text are obtained by directly evaluating these final parameters on the 1,026-material test set that was never used during training or cross-validation.

{\color{black}
\section{Detailed Performance Analysis of Different Models}}


To ensure fair comparison, all models were trained using the same protocol: 100 epochs, batch size of 64, and the Adam optimizer. The learning rates were carefully tuned based on preliminary experiments: ANN and HQCNN use a learning rate of 0.01, while CVNN requires a smaller learning rate of 0.002 due to the complex-valued optimization landscape, where gradients tend to be larger in magnitude. Table S1 summarizes the architectures and parameter counts of the three models.

\begin{table}[htbp]
\centering
\caption{Architecture and parameter count comparison.}
\label{tab:architecture}
\begin{tabular}{l|c|c|c}
\hline
\hline
\textbf{Model} & \textbf{Input dim.} & \textbf{Architecture} & \textbf{Trainable params} \\
\hline
ANN & 54 & 54-30-2 fully connected & $54\times30 + 30\times2 = \mathbf{1,680}$ \\
HQCNN & 64 & Quantum circuit (48 params) + 64-24-2 NN & $48 + 64\times24 + 24\times2 = \mathbf{1,632}$ \\
CVNN & 54 & Complex-valued 54-15-2 & $(54\times15 + 15\times2) \times 2 = \mathbf{1,680}$ \\
\hline
\end{tabular}
\end{table}

The input dimension of HQCNN is 64 because amplitude encoding requires the input vector dimension to match the Hilbert space dimension of the quantum system: a 6-qubit circuit has a $2^6 = 64$-dimensional state space, so the original 54-dimensional elemental fraction vector is padded with 10 zeros before encoding (see Section S1 for details). ANN and CVNN directly use the 54-dimensional input without padding. The slight parameter count difference between HQCNN (1,632) and the baselines (1,680) is less than 3\% and does not affect the validity of comparison.
{\color{black}
We conducted 10-fold cross-validation and evaluated the models on training (7,200 samples), validation (800 samples), and test sets (1,026 samples). The complete results are presented in Figure~\ref{fig:cross_validation}, which shows the performance metrics across all three subsets. These results are discussed in Section~S5.1. The averaged training and validation curves (accuracy and loss) during the cross-validation process are presented in Figure~\ref{fig:cross_validation_curves} and discussed in Section~S5.2.

\begin{figure}[htbp]
\centering
\includegraphics[width=\textwidth]{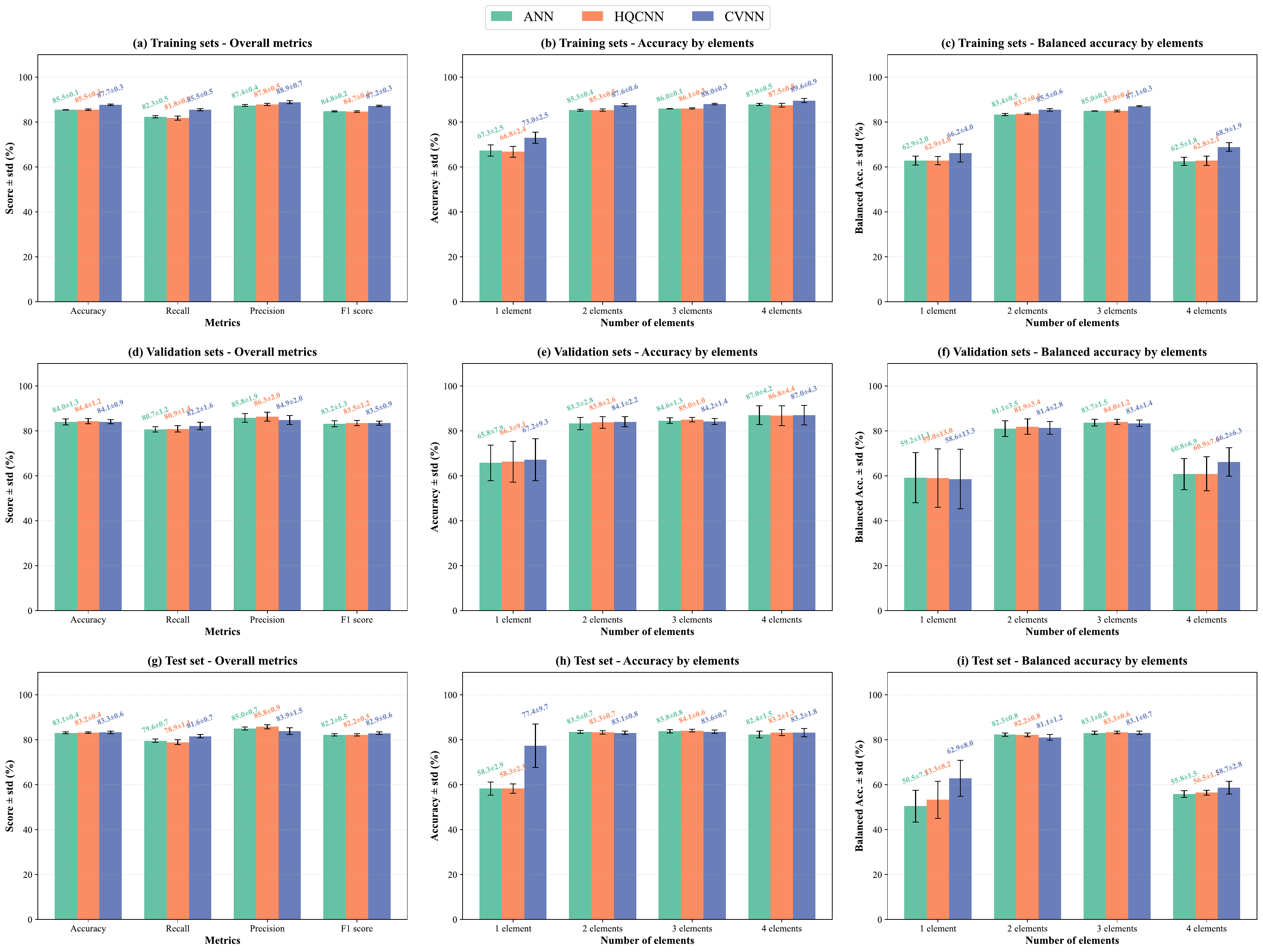}
\caption{Performance comparison of ANN, HQCNN, and CVNN models using 10-fold cross-validation. Results are shown for training sets (top row), validation sets (middle row), and test set (bottom row). For each dataset, three aspects are displayed: overall metrics (Accuracy, Recall, Precision, F1 score), accuracy by elements, and balanced accuracy by elements. Error bars represent standard deviations across the 10 folds. (a) Training sets - Overall metrics. (b) Training sets - Accuracy by elements. (c) Training sets - Balanced accuracy by elements. (d) Validation sets - Overall metrics. (e) Validation sets - Accuracy by elements. (f) Validation sets - Balanced accuracy by elements. (g) Test set - Overall metrics; (h) Test set - Accuracy by elements. (i) Test set - Balanced accuracy by elements.}
\label{fig:cross_validation}
\end{figure}

\subsection{Analysis of Model Performance}
In Figure~\ref{fig:cross_validation}, panels (a)–(c) correspond to training sets, (d)–(f) to validation sets, and (g)–(i) to test sets. Within each group, the first panel shows overall metrics, the second shows accuracy by elements, and the third shows balanced accuracy by elements. The following analysis refers to these panels accordingly.

As an overall performance on the test set, HQCNN achieves an accuracy of 83.2\% and an F1 score of 82.2\%, comparable to ANN (83.1\% accuracy, 82.2\% F1 score). Notably, HQCNN exhibits \textbf{higher precision (85.8\%)} than ANN (85.0\%) and CVNN (83.9\%), indicating the lowest false positive rate in topological material prediction. This suggests that the quantum-native feature extraction helps reduce overprediction of topological labels, which is particularly valuable for high-throughput screening where false positives lead to wasted DFT validation efforts. Meanwhile, CVNN achieves the highest recall on the test set (81.6\%), compared to HQCNN's 78.9\%,  but its precision is lower (83.9\%), indicating that it tends to identify more topological materials at the cost of generating more false positives. The trade-off between the two models is that HQCNN achieves a 1.9 percentage point higher precision than CVNN, while CVNN achieves a 2.7 percentage point higher recall than HQCNN. Given that HQCNN's precision reaches a high absolute value of 85.8\%, HQCNN offers a significant advantage in practical high-throughput screening scenarios.

 For HQCNN, the accuracy decreases from 85.5\% (training) to 84.4\% (validation) to 83.2\% (test), and the F1 score decreases from 84.7\% (training) to 83.5\% (validation) to 82.2\% (test), showing a stable trend overall and demonstrating good generalization. CVNN achieves competitive performance on the training sets, with an accuracy of 87.7\% and an F1 score of 87.2\% — both significantly higher than HQCNN and ANN. However, this advantage does not fully translate to validation and test sets: CVNN's test accuracy and F1 score drop to 83.3\% and 82.9\%., nearly identical to HQCNN's 83.2\% and 82.2\%, respectively. This indicates that CVNN exhibits \textbf{mild overfitting} despite the carefully tuned smaller learning rate (0.002). The complex-valued optimization landscape appears more prone to capturing training set noise, whereas the quantum-native feature extraction in HQCNN provides a form of regularization that improves generalization. The gap between training and test performance is larger for CVNN than for HQCNN in both accuracy (4.4 vs. 2.3 percentage points) and F1 score (4.3 vs. 2.5 percentage points), highlighting the better generalization of the hybrid quantum-classical approach.

Following the analysis framework of Ma et al.,\cite{Ma36} we decomposed model performance based on the number of distinct elements in each compound. Several observations emerge from Figure~\ref{fig:cross_validation}:

\begin{enumerate}

    \item \textbf{Single-element compounds:} With only 213 single-element compounds in the full dataset, model performance shows greater variation and larger standard deviations. Moreover, since no inter-element correlation terms exist for these compounds, this subset does not provide a proper representation of the distinctive capabilities of HQCNN, CVNN, and ANN. Performance on single-element compounds therefore reflects only the model's implementation and optimization, without the contribution of inter-element correlations.

    Nevertheless, to maintain completeness in our comparison, CVNN achieves the highest accuracy (77.4\%) and balanced accuracy (62.9\%) on the test set, suggesting that complex-valued representations may better capture the electronic structure of pure elements. However, the large standard deviations ($\pm9.7\%$ for accuracy, $\pm8.0\%$ for balanced accuracy) indicate that this observed advantage is not statistically robust given the small sample size.

    \item \textbf{Binary and ternary compounds:} With a comparatively larger sample size, binary and ternary compounds provide a statistically reliable basis for performance evaluation. All models demonstrate robust performance on these systems where on the test set, HQCNN achieves accuracies of 83.3\% and 84.1\% for binary and ternary compounds, respectively, with corresponding balanced accuracies of 82.2\% and 83.3\%. These results are comparable to ANN and CVNN, confirming that all models capture the essential chemistry of mid-complexity compounds. However, HQCNN shows a slight advantage on ternary compounds (84.1\% vs. ANN's 83.8\%), suggesting that pairwise interactions—naturally encoded by the quantum measurement process—become more beneficial as chemical complexity increases.

    \item \textbf{Quaternary compounds:} In principle, HQCNN model's that explicitly incorporate correlation effects is expected to perform better as material complexity increases. However, this trend is not fully realized for quaternary compounds, where all models exhibit significantly degraded performance, with balanced accuracies dropping to the 55–59\% range—only marginally above random guessing (50\%). This can be attributed to (i) the much smaller sample size compared to binary and ternary compounds, and (ii) severe label imbalance, with only $\sim$14.7\% of samples labeled as topological.
    
    The uniformly low performance across all models highlights a key limitation of the current dataset rather than the model design, and points to a clear direction for future work: improving predictions for high-complexity materials will require larger, more balanced, and higher-quality datasets for quaternary systems.
\end{enumerate}

\begin{figure}[htbp]
\centering
\includegraphics[width=\textwidth]{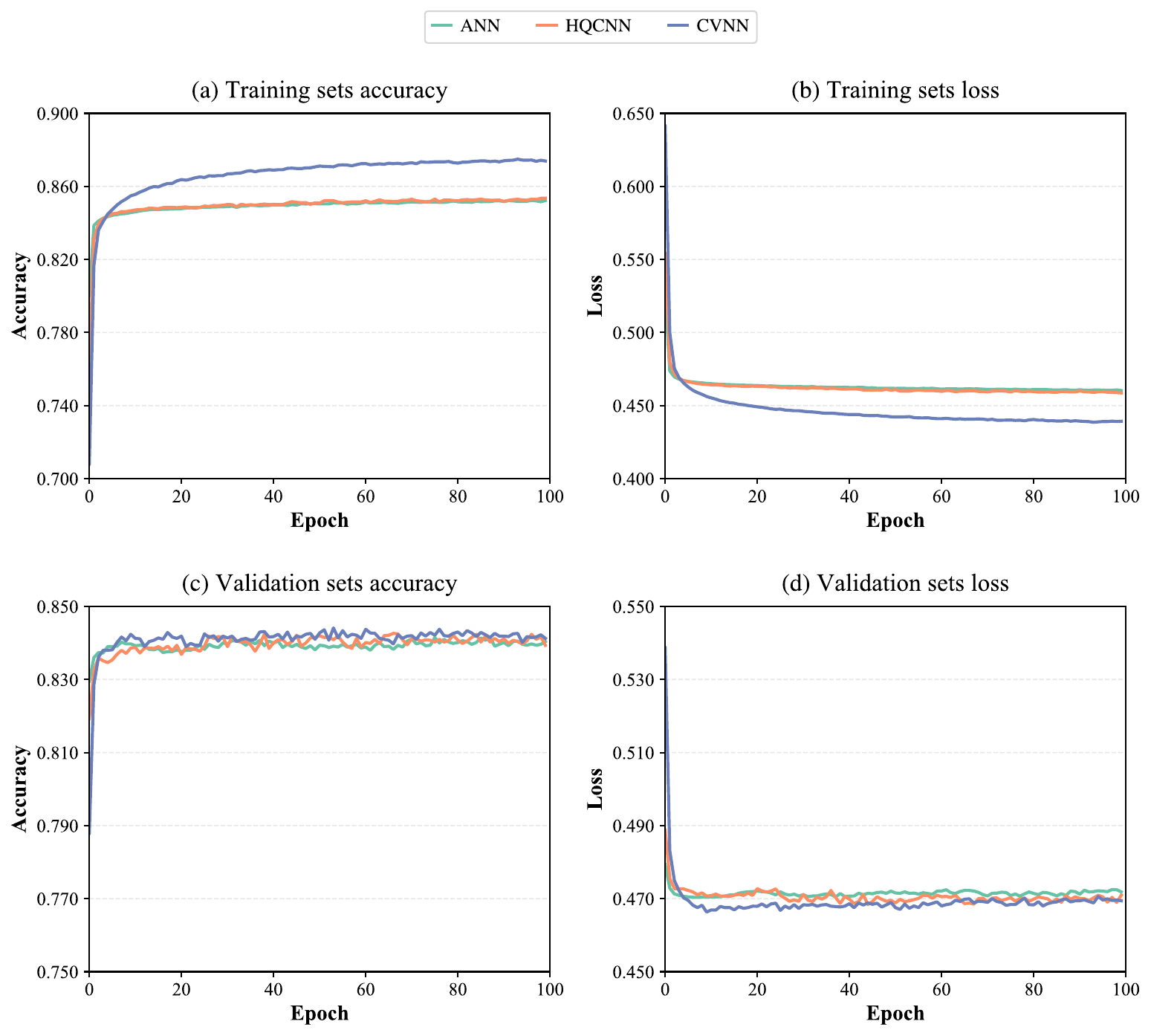}
\caption{{\color{black}Evolution curves of average accuracy and loss on training sets and validation sets for ANN, HQCNN, and CVNN models during ten-fold cross-validation training. (a) Training sets accuracy. (b) Training sets loss. (c) Validation sets accuracy. (d) Validation sets loss.}}
\label{fig:cross_validation_curves}
\end{figure}

\subsection{Training and Validation Curves}
Figure~\ref{fig:cross_validation_curves} further presents the averaged performance evolution curves of the three models during the 10-fold cross-validation training process, where (a) and (b) correspond to the accuracy and loss of training sets, and (c) and (d) correspond to the accuracy and loss of validation sets.

From the training accuracy curves, CVNN maintains the highest training accuracy throughout almost the entire training process, while HQCNN and ANN exhibit similar rising rates. In terms of validation accuracy,  HQCNN and CVNN perform similarly, with both slightly superior to ANN throughout the training process. Similarly, the loss curves for the training sets show that CVNN achieves the largest reduction in training loss, with HQCNN remains slightly lower than ANN overall. In contrast, for the loss curves for the validation sets, CVNN drops to the lowest value in the early stage and then slowly rises, eventually reaching a level comparable to HQCNN by the end of training, while ANN remains the highest. This observation is consistent with the finding in Section S5.2 that CVNN exhibits mild overfitting. In comparison, HQCNN achieves more stable performance on validation loss, demonstrating its robust generalization capability.

\subsection{Summary of HQCNN's Advantages}

From the comprehensive cross-validation analysis, several advantages of the HQCNN model emerge:

\begin{enumerate}
    \item \textbf{Higher precision (85.8\%)} than classical baselines, reducing false positives in topological material prediction—critical for efficient high-throughput screening where each false positive leads to wasted DFT validation resources.
    
    \item \textbf{Comparatively better performance on ternary compounds} (84.1\% accuracy) where pairwise interactions become important, demonstrating that quantum-native feature extraction effectively captures inter-element correlations that are inaccessible to classical heuristics treating elements independently.
    
    \item \textbf{Better generalization} with a smaller train-test performance gap compared to CVNN. The training and validation curves further support this conclusion: HQCNN exhibits more stable validation loss during the training process, implying that the introduction of the quantum circuit provides a certain degree of improvement in model performance.
\end{enumerate}

In summary, the HQCNN model offers a favorable combination of predictive accuracy, interpretability, and generalization for topological material discovery.
}

\section{Confidence Analysis of $g^{Q}(M)$ Score}

In this section, we elaborate on the magnitude of $g^{Q}(M)$ which contains valuable information about the model’s confidence. To quantify this, we introduce a topological confidence score, denoted as $\sigma(B)$, defined for each bin $B$ of $g^{Q}(M)$ values. Here, a bin represents a specific interval on the real axis of possible $g^{Q}(M)$ values.

For any bin $B$, we define:
\begin{equation}
    \sigma(B) =
\frac{
\text{count of topological materials in bin $B$}}{\text{total number of materials in bin $B$}}.
\end{equation}

Thus, $\sigma(B)$ represents the fraction of materials in bin $B$ that are truly topological according to the ground-truth labels.

\begin{figure*}[htbp]
\centering
\includegraphics[width=\linewidth]{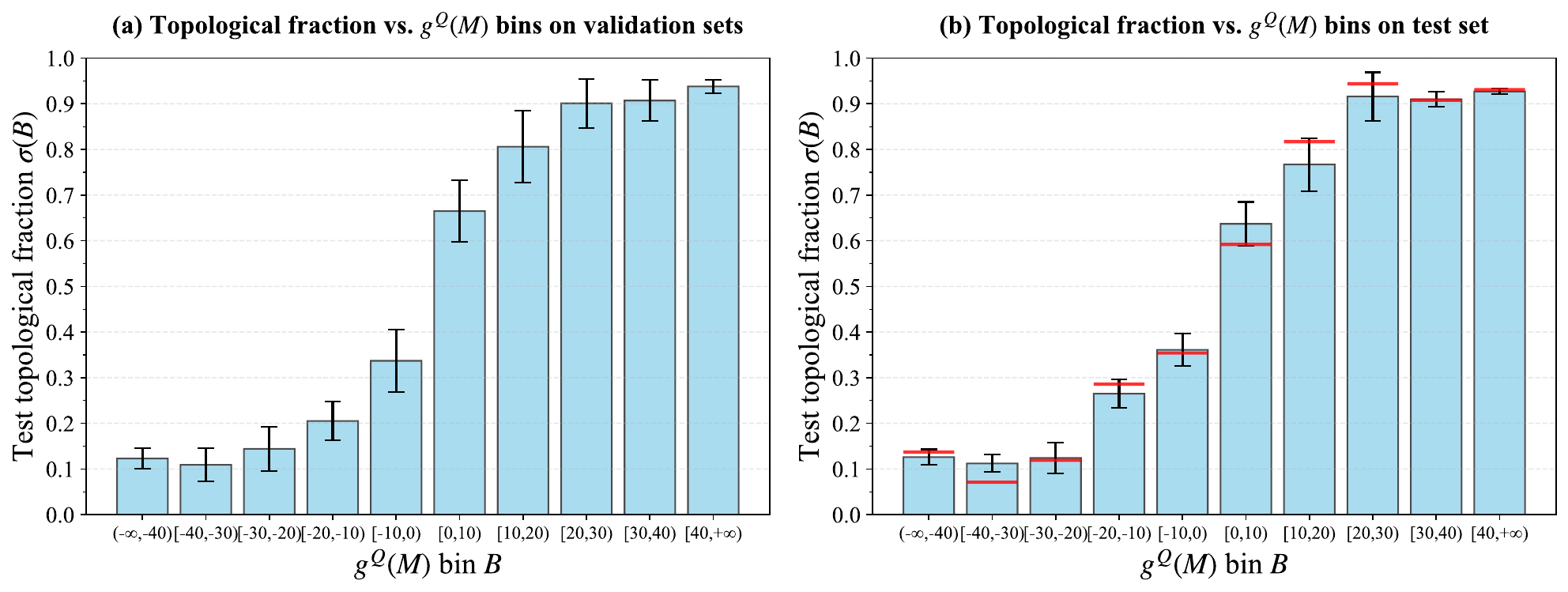}
 \caption{Topological fraction as a function of the $g^{Q}(M)$ bin. \textbf{(a)} Topological fraction vs. $g^{Q}(M)$ bins evaluated on validation sets. \textbf{(b)} Topological fraction vs. $g^{Q}(M)$ bins on test set. For each bin $B$, the bar height indicates the mean topological fraction $\sigma(B)$, and the error bars represent $\pm$ one standard deviation. These means and standard deviations are computed from the ten cross-validation folds. In panel (b), the red line shows the topological fraction obtained on the final test set using the aggregated parameter values employed to compute $g^{Q}(M)$.}
\label{fig:binb}
\end{figure*}

To analyze the behavior of this score, we compute the mean and standard deviation of $\sigma(B)$ across all folds in our 10-fold cross-validation. The resulting trend is highly informative as we demonstrate it for our validation and test sets (Figure~\ref{fig:binb}). When $g^{Q}(M) < -20$, the topological score remains low and stable, indicating that almost all materials in this range are trivial. In the intermediate region $-20 < g^{Q}(M) < 20$, the score increases steadily, reflecting growing uncertainty in classification. Once $g^{Q}(M) > 20$, the score becomes both high and stable, revealing that materials in this range are overwhelmingly topological.

This histogram can also be interpreted directly in terms of classification reliability. For the first five bins (negative $g^Q(M)$  values), the model predicts materials to be trivial. In these bins, $\sigma(B)$ therefore represents the misclassification rate, and the accuracy equals $1 - \sigma(B)$. For the remaining positive-valued bins, the model predicts non-trivial topology, so $\sigma(B)$ directly gives the correct classification rate in that region.

Across both sides of the decision boundary, the classification accuracy improves as $|g^{Q}(M)|$ increases. This accuracy eventually reaches a plateau once $|g^{Q}(M)| \approx 20$, beyond which additional increases in magnitude do not significantly change the reliability. This saturation provides strong, quantitative evidence that $|g^{Q}(M)| \ge 20$ serves as a natural and robust threshold for high-confidence classification, enabling reliable screening in large-scale materials discovery. Focusing specifically on topological predictions, we evaluate the final averaged parameters on the independent test set and find that for materials predicted to be topological ($g^{Q}(M) \ge 20$), the classification accuracy reaches 92.8\%, compared to the overall test set accuracy of 83.5\%. This substantial improvement confirms the practical utility of this threshold for identifying high-confidence topological candidates.

Our approach to determining the confidence threshold follows the empirical methodology of Ma et al.~\cite{Ma36}, who observed a similar saturation behavior for $g(M) \ge 1$ in their topogivity model when identifying topological materials. The difference in numerical threshold (20 vs.~1) reflects the different scales of the diagnostic functions, which arise from differences in model architecture and parameterization. In both cases, the threshold is empirically determined from the saturation behavior and serves as a practical guide for high-confidence screening, rather than a rigorously optimized parameter. We note that this threshold may depend on the dataset and model architecture; when applying the method to new datasets, recalibration may be necessary.

\clearpage
\bibliography{AMI_revise}